\documentclass[]{aastex63}

\received{XXXX XX, 2023}
\revised{XXXX XX, 2023}
\accepted{XXXX YY, 2023}
\submitjournal{AJ}
\shorttitle{The open clusters King 6 and NGC 1605}
\shortauthors{Gokmen et al.}
\usepackage{xcolor}
\usepackage{rotating}
\usepackage{booktabs}  
\begin{document}

\title{CCD UBV and Gaia DR3 Analyses of Open Clusters King 6 and NGC 1605}

\correspondingauthor{Sevinc Gokmen}
\email{sgokmen2021@fau.edu}

\author[0000-0002-0108-4782]{Sevinc Gokmen}
\affiliation{Department of Physics, Florida Atlantic University, Boca Raton, Florida 33431, USA}

\author[0000-0003-1883-6255]{Zeki Eker}
\affiliation{Akdeniz University, Faculty of Sciences, Department of Space Sciences and 
Technologies, 07058, Antalya, Turkey}

\author[0000-0002-5657-6194]{Talar Yontan}
\affiliation{Istanbul University, Faculty of Science, Department of Astronomy and Space Sciences, 34119, Beyaz\i t, Istanbul, Turkey}

\author[0000-0003-3510-1509]{Sel\c{c}uk Bilir}
\affiliation{Istanbul University, Faculty of Science, Department of Astronomy and Space Sciences, 34119, Beyaz\i t, Istanbul, Turkey}

\author[0000-0002-0688-1983]{Tansel Ak}
\affiliation{Istanbul University, Faculty of Science, Department of Astronomy and Space Sciences, 34119, Beyaz\i t, Istanbul, Turkey}

\author[0000-0002-0912-6019]{Serap Ak}
\affiliation{Istanbul University, Faculty of Science, Department of Astronomy and Space Sciences, 34119, Beyaz\i t, Istanbul, Turkey}

\author[0000-0001-9445-4588]{Timothy Banks}
\affiliation{Nielsen, 675 6th Ave, New York, NY 10011, USA}
\affiliation{Dept.\ of Physical Science \& Engineering, Harper College, 1200 W Algonquin Rd, IL 60067, USA}

\author[0000-0001-6708-4374]{Ata Sarajedini}
\affiliation{Department of Physics, Florida Atlantic University, Boca Raton, Florida 33431, USA} 
\begin{abstract}
\noindent

A detailed analysis of ground-based CCD {\it UBV} photometry and space-based {\it Gaia} Data Release 3 (DR3) data for the open clusters King 6 and NGC 1605 was performed. Using the {\sc pyUPMASK} algorithm on {\it Gaia} astrometric data to estimate cluster membership probabilities, we have identified 112 stars in King 6 and 160 stars in NGC 1605 as the statistically most likely members of each cluster. We calculated reddening and metallicity separately using {\it UBV} two-color diagrams to estimate parameter values via independent methods. The color excess $E(B-V)$ and photometric metallicity [Fe/H] for King 6 are $0.515 \pm 0.030$ mag and $0.02 \pm 0.20$ dex, respectively. For NGC 1605, they are $0.840 \pm 0.054$ mag and $0.01 \pm 0.20$ dex. With reddening and metallicity kept constant, we have estimated the distances and cluster ages by fitting PARSEC isochrones to color-magnitude diagrams based on the {\it Gaia} and {\it UBV} data. Photometric distances are 723 $\pm$ 34 pc for King 6 and 3054 $\pm$ 243 pc for NGC 1605. The cluster ages are $200 \pm 20$ Myr and $400 \pm 50$ Myr for King 6 and NGC 1605, respectively. Mass function slopes were found to be 1.29 $\pm$ 0.18 and 1.63 $\pm$ 0.36 for King 6 and NGC 1605, respectively. These values are in good agreement with the value of \citet{Salpeter55}. The relaxation times were estimated as 5.8 Myr for King 6  and 60 Myr for NGC 1605. This indicates that both clusters are dynamically relaxed since these times are less than the estimated cluster ages. Galactic orbit analysis shows that both clusters formed outside the solar circle and are members of the young thin-disc population.

\end{abstract}

\section{Introduction}
\label{sec:introduction}

The study of open clusters (OCs) gives insights into both stellar and galactic evolution.  The stars making up a cluster are formed at essentially the same time and initially share the same kinematic and positional behaviour. While the distances, ages, and chemical compositions of the cluster stars are similar, their masses differ \citep{Maurya20}. In studying a cluster, a number of parameters can be taken as fixed for all of the stars, such as distance and age. This means that differences in the apparent magnitudes of the cluster's stars will primarily be due to mass, making OCs very useful in such work. Beyond investigation of the component stars of a cluster, clusters themselves can act as tracers of the chemical enrichment of the Galaxy with time as generations of stars return their constituent atoms back into the interstellar medium and ultimately into later generations of stars \citep{Mckee07, Cantat-Gaudin20, Hou21, He21} as they form new clusters.    

In this paper, we investigate the open clusters King~6 and NGC~1605, which are located in the second Galactic quadrant. Both clusters are very close to the Galactic plane. Moving close to this plane results in increased field contamination. Such contamination makes identification of stars having cluster membership, as distinct from being in the surrounding field and therefore unrelated to the cluster, more difficult. This paper will discuss the statistical, careful removal of such field contamination. This is necessary for the precise determination of cluster parameters. This study is part of a wider project that investigates the detailed properties of open clusters in the Galaxy \citep[cf.][]{Yontan15, Yontan19, Yontan21, Yontan22, Yontan23b, Yontan23a}. The paper aims to estimate the main parameters of King~6 and NGC~1605 using space-based {\it Gaia} DR3 data \citep{Gaia23} and ground-based {\it UBV} photometric data. 

Turning to the actual clusters themselves:
\begin{itemize}
    \item{King~6 ($\alpha=03^{\rm h} 27^{\rm m} 56^{\rm s}$, $\delta= +56^{\rm o} 26^{\rm '} 39^{\rm''}$, $l=143\,^{\rm o}\!\!.\,3444$, $b=-0\,^{\rm o}\!\!.\,0949$) was classified by \citet{Ruprecht66} as Trumpler type IV 2p with a weak central stellar concentration. \cite{Ann02} presented CCD {\em UBVI} photometry, color-magnitude diagrams, and main sequence isochrone fitting for the cluster. They noted strong contamination of the main sequence fainter than $V \sim 18$ mag by field stars, and that the color-color diagram could not be fitted by a theoretical zero-age main sequence (ZAMS) with a single reddening value. Stars fainter than $V \sim 16$ mag appeared to have a lesser $E(B-V)$ of 0.4 mag, with stars brighter than $V \sim 13$ mag having an extinction of 0.6 mag. A mean value of $0.5 \pm 0.1$ mag was adopted by \citet{Ann02}, leading to what they noted as a poor isochrone fit. The cluster age was estimated a $\log{t} = 8.4 \pm 0.1$, [Fe/H] as 0.46 dex, and the apparent distance modulus as $11.25 \pm 0.71$ mag. They also commented that there could be a binary sequence above the ZAMS. \cite{Maciejewski07} collected CCD {\em BV} photometry of the cluster. They fit a \cite{King62} profile with a core radius of $3.6 \pm 0.4$ arcminutes, densities $f_{\rm 0} =1.55 \pm 0.09$ and $f_{\rm bg} = 0.62 \pm 0.04$ stars per square arcminute, and a limiting radius as 10.9 arcminutes. Here, core radius refers to the radial distance from the cluster center within which the density drops to the half of the central density, $f_0$ refers to the number density of stars in the center of the cluster, and $f_{\rm bg}$ refers to the number density of field stars. The log age of King~6 was estimated as $8.4$, the distance modulus as $11.17^{+0.55}_{-0.47}$ mag, and $E(B-V) = 0.53^{+0.12}_{-0.11}$ mag. The `slope' of the mass function was estimated as $1.74 \pm 0.39$ with no statistically different slopes for the core ($1.44 \pm 0.32$) and halo ($1.58 \pm 0.47$), and so not suggesting the presence of mass segregation within the cluster.} 
    \item{NGC 1605 ($\alpha=04^{\rm h} 34^{\rm m} 59^{\rm s}$, $\delta= +45^{\rm o} 16^{\rm '} 08^{\rm''}$, $l=158\,^{\rm o}\!\!.\,5860$, $b=-01\,^{\rm o}\!\!.\,5673$) was classified by \citet{Ruprecht66} as a Trumpler type III 1m of medium richness. \cite{Fang_1970} applied {\it RGU} photometry to construct color-magnitude diagrams, and commented on the lack of red giants in the cluster. A distance of 2750 pc was estimated. \cite{Sujatha03} presented CCD {\it UBVRI} photometric observations, finding a $E(B-V)$ of 0.14 mag and a distance of 1148 pc. \cite{Camargo21} noted ``an unusual morphology with a sparse stellar distribution and a double core in close angular proximity'', i.e. a binary cluster. One of the cores was estimated as being substantially older, at an age of $2 \pm 0.2$ Gyr compared to the $600 \pm 100$ Myr of the other core. \cite{Camargo21} argued for tidal capture being the origin of the binary cluster. Joint distance was estimated as $2.6\pm 0.4$ kpc.}
\end{itemize}
Both clusters are also explored in general surveys. See Table~\ref{tab:literature} for summaries of results from the literature for both clusters.

The astrophysical parameters of OCs are commonly estimated simultaneously from isochrones fitted to the observed color–magnitude diagrams (CMDs) \citep{Angelo21, Bisht22} and Bayesian statistics \citep{vonHippel06, Bossini19}. Large uncertainties in the derived reddening, metallicity and hence age are driven by degeneracies between parameters in such simultaneous statistical solutions based on the comparison of stellar isochrones with photometric observations \citep{Janes14}. To break the reddening-age degeneracy in these simultaneous solutions, several approaches have been proposed. Using the available wavelength range, preferably including a near-infrared band, is the main idea behind most of these advances \citep{Bilir06, Bilir10, deMeulenaer13}. According to \citet{Anders04}, when near-infrared observations are not available, one of the most suitable photometric band combinations for reliable parameter determination is based on {\em UBVRI} photometry. In addition traditional and reliable methods developed for determining reddening and metallicity can be used to constrain these parameters \citep{Ak16, Bilir16, Banks20}. In this current study, reddening and metallicity of the clusters were obtained via independent methods from the {\em UBV} photometric data. Hence, the parameter degeneracy between reddening and age values was minimized.

In this paper, we determine the cluster membership probabilities for stars in the general line of sight of the clusters, mean distances and proper-motion components of King 6 and NGC 1605. We present structural and fundamental parameters, luminosity and mass functions, the dynamical states of mass segregation, and kinematic and Galactic orbital parameters of both clusters.

\begin{table*}
\setlength{\tabcolsep}{1.5pt}
\renewcommand{\arraystretch}{0.5}
  \centering
  \caption{Summary of results from the literature for the King 6 and NGC 1605 OCs. Columns are color excess ($E(B-V$)), distance ($d$), iron abundance ([Fe/H]), age ($t$), proper-motion components ($\langle\mu_{\alpha}\cos\delta\rangle$, $\langle\mu_{\delta}\rangle$), and radial velocity ($V_{\rm R}$). `Ref' indicates the source of the data, according to the list of papers below the table.}
  \begin{tabular}{cccccccl}
    \toprule
    \multicolumn{8}{c}{King 6}\\
        \toprule
$E(B-V)$ & $d$ & [Fe/H] & $t$ &  $\langle\mu_{\alpha}\cos\delta\rangle$ &  $\langle\mu_{\delta}\rangle$ & $V_{\rm R}$ & Ref \\
(mag) &  (pc)  & (dex) & (Myr) & (mas yr$^{-1}$) & (mas yr$^{-1}$) & (km s$^{-1})$ &      \\
        \bottomrule
        \toprule
0.50$\pm$0.10   & 870$\pm$180 & 0.46             & 250$\pm$50 & ---              & ---                & ---               & (01) \\
0.58            & 450         & ---              & 260        & $-$0.77$\pm$0.78 & $-$2.01$\pm$1.23   & ---               & (02) \\
0.53$\pm$0.12   & 800$\pm$270 & ---              & 250        & ---              & ---                & ---               & (03) \\
0.30            & 632         & ---              & 950        & +2.45            & $-$1.82            & ---               & (04) \\
0.50            & 871         & ---              & 250        & ---              & ---                & ---               & (05) \\
0.624           & 763         & ---              & ~~9        & +2.700$\pm$0.700 & $-$3.006$\pm$0.675 & ---               & (06) \\
 ---            & 727         & ---              & ---        & ---              & ---                & $-$21.88$\pm$2.56 & (07) \\  
 ---            & 727$\pm$53  & ---              & ---        & +3.864$\pm$0.016 & $-$1.814$\pm$0.017 & ---               & (08) \\
 ---            & 744$\pm$39  & ---              & 47$\pm$3   & +3.871$\pm$0.274 & $-$1.821$\pm$0.290 & ---               & (09) \\
0.357$\pm$0.015 & 807$\pm$18  & 0.46             & 380$\pm$60 & ---              & ---                & ---               & (10) \\
0.357$\pm$0.015 & 727         & $-$0.09$\pm$0.05 & ---        & +3.864$\pm$0.016 & $-$1.814$\pm$0.017 & $-$26.78$\pm$8.17 & (11) \\
0.43            & 751         & ---              & 200        & +3.864$\pm$0.016 & $-$1.814$\pm$0.017 & ---               & (12) \\
0.595$\pm$0.021 & 704$\pm$14  & +0.15$\pm$0.07   & 195$\pm$90 & +3.880$\pm$0.250 & $-$1.825$\pm$0.257 & $-$22.14$\pm$7.34 & (13) \\
 ---            & 738         & ---              & 210        & ---              & ---                & $-$22.09$\pm$3.48 & (14) \\
0.595$\pm$0.062 & 707$\pm$1   & ---              & 143$\pm$66 &  +3.825$\pm$0.011 &  $-$1.910$\pm$0.010 &  $-$21.89$\pm$1.90 & (15) \\
0.515$\pm$0.030 & 723$\pm$34  & +0.02$\pm$ 0.20   & 200$\pm$20 & +3.833$\pm$0.034 & $-$1.906$\pm$0.032 & $-$23.40$\pm$3.26 & (16) \\

  \bottomrule
    \multicolumn{8}{c}{NGC 1605}\\
    \bottomrule
    \toprule
 ---            & 2800         & ---              &  65   & ---                & ---                & ---                & (17) \\
1.19            & ---          & ---              &  125        & ---                & ---                & ---                & (18) \\
0.14            & 1148         & ---              & 5000        & ---                & ---                & ---                & (19) \\  
0.97            & 2600         & ---              &  40         & $-$0.47            & $-$3.08            & ---                & (04) \\
0.97            & 2559         & ---              &  40         & ---                & ---                & ---                & (05) \\
0.591           & 1996         & ---              & 420         & $-$0.529$\pm$0.347 & $-3$.985$\pm$0.401 & ---                & (06) \\
 ---            & 3083         & ---              & ---         & ---                & ---                & $-$1.15$\pm$0.12   & (07) \\ 
 ---            & 3083$\pm$1050& ---              & ---         & +0.990$\pm$0.017   & $-1$.977$\pm$0.012 & ---                & (08) \\
 ---            & 3322$\pm$519 & ---              & 676$\pm$41  & +0.924$\pm$0.267   & $-1$.911$\pm$0.318 & ---                & (09) \\
0.970           & 3083         & $-$0.04$\pm$0.37 & ---         & +0.990$\pm$0.151   & $-1$.977$\pm$0.104 & $-$12.03$\pm$17.42 & (11) \\
0.713           & 3073         & ---              & 190         & +0.990$\pm$0.151   & $-1$.977$\pm$0.104 & ---                & (12) \\
---             & 3083$\pm$49  & ---              & ---         & +0.990$\pm$0.017   & $-1$.977$\pm$0.012 & ---                & (20) \\
0.817$\pm$0.020 & 2449$\pm$98  & $-$0.08$\pm$0.07 & 380$\pm$25  & +1.002$\pm$0.192   & $-$2.000$\pm$0.139 & ---                & (13) \\
---             & 2600$\pm$400 & ---              & 600$\pm$100 & ---                & ---                & ---                & (21) \\
---             & 3073         & ---              & 190         & +0.935$\pm$0.097   & $-1$.989$\pm$0.052 & ---                & (22) \\
---             & 2887         & ---              & 260         & ---                & ---                & $-1.15 \pm$0.12    & (14) \\
0.831$\pm$0.060 & 2735$\pm$31   &  ---         & 215$\pm$93 &  +0.948$\pm$0.005 & $-$2.002$\pm$0.007 &  $-$1.55$\pm$0.52 &  (15) \\
0.840$\pm$0.054 & 3054$\pm$243 & +0.01$\pm$0.20   & 400$\pm$50  & +0.928$\pm$0.104   & $-$1.997$\pm$0.082 & $-$15.27$\pm$1.35  & (16) \\
    \bottomrule
    \toprule
    \end{tabular}%
    \\
\raggedright
{\scriptsize
(01)~\citet{Ann02}, (02)~\citet{Kharchenko05}, (03)~\citet{Maciejewski07}, (04)~\citet{Kharchenko12}, (05)~\citet{Sampedro17}, (06)~\citet{Loktin17}, (07)~\citet{Soubiran18}, (08)~\citet{Cantat-Gaudin18}, (09)~\citet{Liu19}, (10)~\citet{Bossini19}, (11)~\citet{Zhong20}, (12)~\citet{Cantat-Gaudin20}, 
(13)~\citet{Dias21}, (14)~\citet{Tarricq21}, (15)~\citet{Hunt23}, (16)~This study, (17)~\citet{Tignanelli90}, (18)~\citet{Ahumada95}, (19)~\citet{Sujatha03}, 
(20)~\citet{Cantat-Gaudin20A}, (21)~\citet {Camargo21}, and (22)~\citet {Poggio21}.
}
  \label{tab:literature}%
\end{table*}%

\begin{figure*}
\centering
\includegraphics[scale=0.25, angle=0]{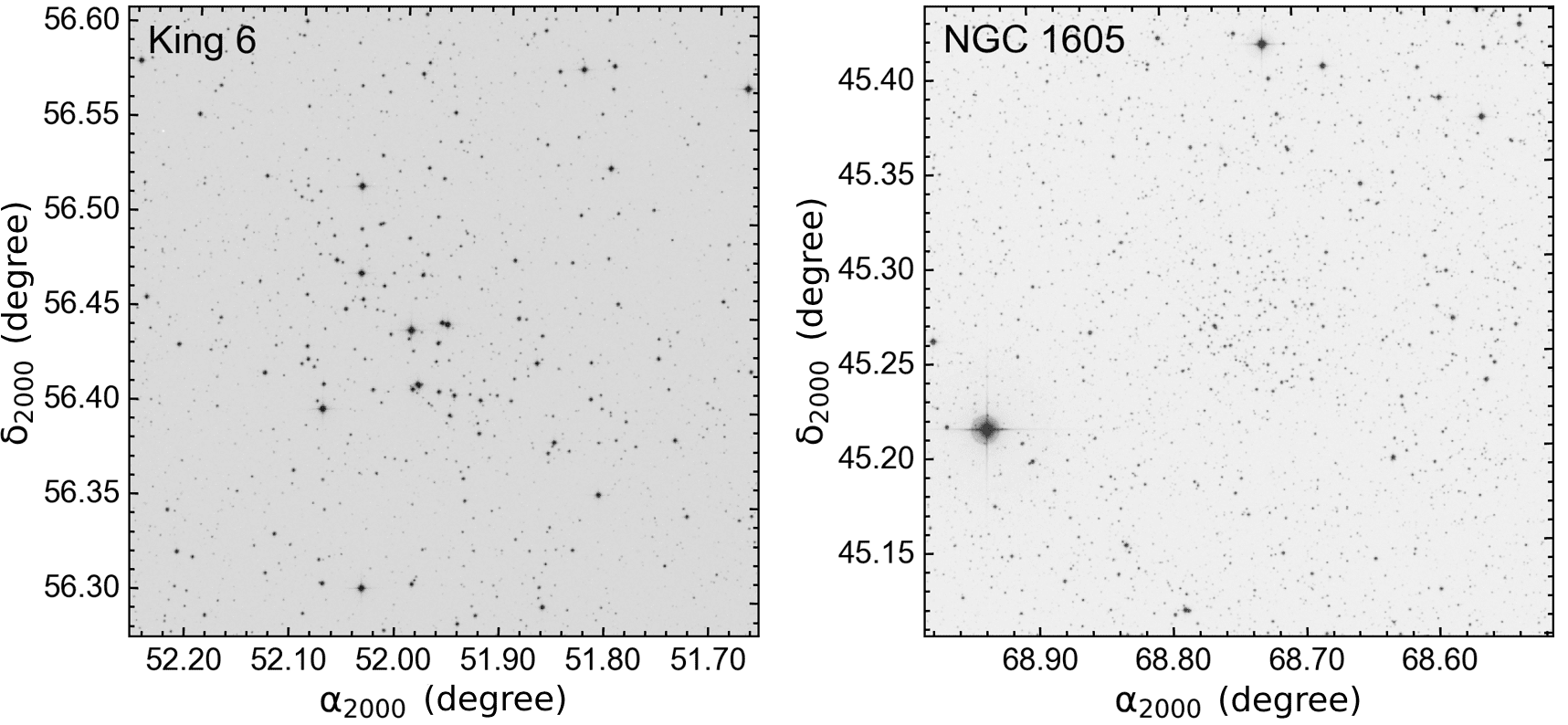}
\caption{$V$-band identification maps for King 6 (left panel) and NGC 1605 (right panel). The field of view of the charts is $21'\!.5 \times 21'\!.5$. North is towards up and East is leftwards.} 
\label{fig:charts}
\end {figure*}


\begin{table}[t]
\setlength{\tabcolsep}{4.5pt}
\renewcommand{\arraystretch}{0.6}
  \centering
  \caption{Log of observations for the King 6 and NGC 1605 OCs. Columns give the cluster name, date of observation (DMY), and information about the exposures by the three photometric filters. In the filter columns, there are two numbers separated by the multiplication symbol.  The first numbers in each such pair give the exposure times (in seconds), while the second indicates the number of exposures ($N$).}
  \medskip
    \begin{tabular}{cclll}
    \hline
            &  & \multicolumn{3}{c}{Filter/Exp, Time (s) $\times N$}        \\
    \hline
    Cluster & Obs. Date  & $U$            & $B$          & $V$              \\
    \hline
King 6   & 30-09-2019 &  $200 \times 3$  &  $20 \times 3$ &   $12 \times 5$ \\
         &            &  $1800 \times 2$ & $600 \times 3$ &  $600 \times 3$ \\
NGC 1605 & 24-09-2014 &  $300 \times 6$  & $120 \times 7$ &   $60 \times 6$ \\
    \hline
    \end{tabular}%
  \label{tab:exposures}%
\end{table}%

\section{Observation and Data Reduction}
\subsection{CCD UBV Observations}
Broadband {\it UBV} CCD photometric observations of the two clusters were performed using the $f/10$ Ritchey-Chr\'etien focus of the 1m T100 telescope. The T100 is installed at the  T\"UB\.ITAK National Observatory (TUG)\footnote{www.tug.tubitak.gov.tr} in Turkey. The CCD is a back-illuminated device with 4k$\times$4k pixels. Each pixel of the CCD detector is 15 $\mu$ across. This corresponds to a $0\,''\!\!.\,31$ pixel$^{-1}$ square on the sky. The CCD therefore has a field of view of $21\,'\!\!.5 \times 21\,'\!\!.5$. The gain is 0.55 e$^{-}$/ADU and the readout noise 4.19~e$^{-}$ (100 KHz).

The observations of King 6 and NGC 1605 were carried out on 30 September 2019 and 24 September 2014, respectively. Bias frames and {\it UBV} flat field frames were taken at the beginning of each night. Sets of long and short exposures were collected for the two clusters so as to avoid saturating the brighter stars and also allow the detection of fainter stars. Identification charts are presented in Figure~\ref{fig:charts}. The log of the observations is given as Table~\ref{tab:exposures}. To obtain the equation of extinction and the associated transformation coefficients, 16 Landolt fields were observed. These covered a total of 156 standard stars. The airmass ranges from 1.13 to 1.91 across these standards. The observed \citet{Landolt09} fields are listed in Table~\ref{tab:standard_stars}.

\subsection{Reductions and Photometric Calibrations}
Bias subtraction and flat fielding were carried out using standard IRAF\footnote{IRAF is distributed by the National Optical Astronomy Observatories} packages. Instrumental magnitudes of the stars were measured using the IRAF aperture photometry packages \citep{Landolt09}. The photometric extinction and transformation coefficients were obtained by employing multiple linear regression fits to these calculated instrumental magnitudes. The resulting coefficients are listed in Table~\ref{tab:coefficients} for the two observing nights. Applying PyRAF\footnote{PyRAF is a product of the Space Telescope Science Institute, which is operated by AURA for NASA} and astrometry.net\footnote{http://astrometry.net}, astrometric calibrations were performed for all cluster frames. Coordinates of detected stars in all images of the cluster fields were aligned and combined in each filter for different exposures. This improved the signal-to-noise ratio for fainter stars and made it possible to measure magnitudes of brighter stars that were saturated in long exposures. The photometry of detected objects in the cluster regions was performed using the Source Extractor (SExTractor) and PSF Extractor (PSFEx) routines \citep{Bertin96}. Aperture corrections were then applied to these magnitudes. Finally, the instrumental magnitudes were transformed to standard magnitudes in the $U\!BV$ filters using expressions from \citet{Janes11}, namely:   

\begin{table}[t]
\setlength{\tabcolsep}{3pt}
\renewcommand{\arraystretch}{0.6}
  \centering
  \caption{Selected \citet{Landolt09} standard star fields. The columns denote the observation date (day-month-year), star field name from Landolt, the number of standard stars ($N_{\rm st}$) observed in a given field, the number of observations for each field ($N_{\rm obs}$), and the airmass range the fields were observed over ($X$).}
\medskip
    \begin{tabular}{llccc}
    \hline
Date	   & Star Field	& $N_{\rm st}$ & $N_{\rm obs}$ & $X$          \\
\hline
      	   & GD 246     &  4 	       & 1	        &              \\
      	   & SA92SF2    & 15 	       & 2	        &              \\
      	   & SA92SF3    &  6	       & 1	        &              \\
              & SA93	     &  4           & 3	         &              \\
24-09-2014    & SA94	   &  2	          & 1	       & 1.13 -- 1.74 \\
              & SA95SF2    &  9	        & 1	         &              \\
	       & SA96	    &  2	       & 1	        &              \\
	       & SA98	    & 19	       & 2	        &              \\
              & SA112	     &  2	        & 1	         &              \\
	       & SA113	    &  1	       & 1	        &              \\
	       & SA114	    &  2	       & 1	        &              \\
\hline	      
              & SA92SF2    & 15           & 1	         &              \\
	       & SA93       &  4	       & 1	        &              \\
	       & SA94	    &  2 	       & 1	        &              \\
              & SA96       &  2	        & 1	         &              \\
              & SA98	     & 19	        & 1	         &              \\
	       & SA108      &  2	       & 1	        &              \\
30-09-2019	  & SA109	   &  2	          & 1	       & 1.23 -- 1.91 \\
	       & SA109SF2   &  3	       & 1	        &              \\
	       & SA110SF2   & 10	       & 1	        &              \\
	       & SA111	    &  5	       & 1	        &              \\	
	       & SA112	    &  6	       & 2	        &              \\
 	       & SA113	    & 15	       & 1	        &              \\       
	       & SA114	    &  5	       & 2	        &              \\	       
    \hline
    \end{tabular}%
  \label{tab:standard_stars}%
\end{table}%

\begin{table*}
\renewcommand{\arraystretch}{0.6}
  \centering
  \caption{Transformation and extinction coefficients obtained for the two observation nights: $k$ and $k'$ are the primary and secondary extinction coefficients, while $\alpha$ and $C$ are the transformation coefficients.} 
  \medskip
    \begin{tabular}{lccccc}
    \hline
Obs. Date & Filter/color index &  $k$   & $k'$             & $\alpha$          & $C$             \\
    \hline
30.09.2019 & $U$     & 0.413$\pm$0.069  & +0.014$\pm$0.094 & ---               & ---             \\
           & $B$     & 0.285$\pm$0.061  & +0.029$\pm$0.069 & 0.852$\pm$0.105   & 1.850$\pm$0.092 \\
           & $V$     & 0.182$\pm$0.002  & ---              & ---               & ---             \\
           & $U-B$   &  ---             & ---              & 0.801$\pm$0.137   & 4.338$\pm$0.102 \\
           & $B-V$   &  ---             & ---              & 0.066$\pm$0.008   & 1.853$\pm$0.033 \\
24.09.2014 & $U$     &  0.330$\pm$0.071 & -0.038$\pm$0.136 & ---               & ---             \\ 
           & $B$     &  0.263$\pm$0.070 & -0.065$\pm$0.083 & 1.004$\pm$0.124   & 0.944$\pm$0.104 \\
           & $V$     & 0.130$\pm$0.010  & ---              & ---               & ---             \\
           & $U-B$   &  ---             & ---              & 0.929$\pm$0.196   & 3.487$\pm$0.102 \\
           & $B-V$   &  ---             & ---              & 0.086$\pm$0.020   & 1.006$\pm$0.030 \\
\hline
    \end{tabular}%
  \label{tab:coefficients}%
\end{table*}%

\begin{equation}
V = v - \alpha_{bv}(B-V)-k_vX _v- C_{bv} \\
\end{equation}

\begin{equation}
B-V = \frac{(b-v)-(k_b-k_v)X_{bv}-(C_b-C_{bv})}{\alpha_b+k'_bX_b-\alpha_{bv}} \\
\end{equation}

\begin{eqnarray}
U-B = \frac{(u-b)-(1-\alpha_b-k'_bX_b)(B-V)}{\alpha_{ub}+k'_uX_u}-\frac{(k_u-k_b)X_{ub}-(C_{ub}-C_b)}{\alpha_{ub}+k'_uX_u} 
\end{eqnarray}
In equations (1-3) $U\!$, $B$, and $V$ are the magnitudes in the standard photometric system. $u$, $b$, and $v$ represent the instrumental magnitudes. $X$ is the airmass. $k$ and $k'$ are primary and secondary extinction coefficients. $\alpha$ and $C$ indicate transformation coefficients to the standard system.

\newpage
\section{Data Analyses}
\subsection{UBV and Complementary Gaia DR3 Data}

Final {\it UBV} photometric catalogs comprise 884 and 2474 stars with magnitudes brighter than $V=22$ mag for King 6 and NGC 1605, respectively. These optical catalogues were cross-matched with {\it Gaia} DR3 photometric and astrometric data, which resulted in the final catalogues containing IDs, positions ($\alpha, \delta$), $V$ apparent magnitudes and $U-B$, $B-V$ color indices, {\it Gaia} DR3 $G$ magnitudes and $G_{\rm BP}-G_{\rm RP}$ color indices, $\mu_{\alpha}\cos\delta, \mu_{\delta}$ proper-motion components along with $\varpi$ trigonometric parallaxes, and $P$ membership probabilities as estimated by this study (see Table~\ref{tab:all_cat}). Both catalogues are available electronically. The internal errors resulting from the PSF (point-spread function) fitting procedure were adopted as the photometric accuracies for the $V$ magnitude, and $U-B$, $B-V$ color indices.  In Table~\ref{tab:phiotometric_errors} we list the mean photometric errors in the Johnson and {\it Gaia} DR3 filters as a function of the $V$ magnitude. It can be seen from the table that stars brighter than $V=22$ mag have mean errors lower than 0.04 mag in $V$ magnitude, and lower than 0.16 mag in $U-B$ and $B-V$ for King 6, while for NGC 1605, these measurements reach up to 0.05 mag in $V$ magnitude, 0.24 and 0.10 mag in $U-B$ and $B-V$ colors, respectively. The {\it Gaia} DR3-based mean errors are lower than 0.08 mag for the two clusters.

\begin{figure*}
\centering
\includegraphics[scale=0.6, angle=0]{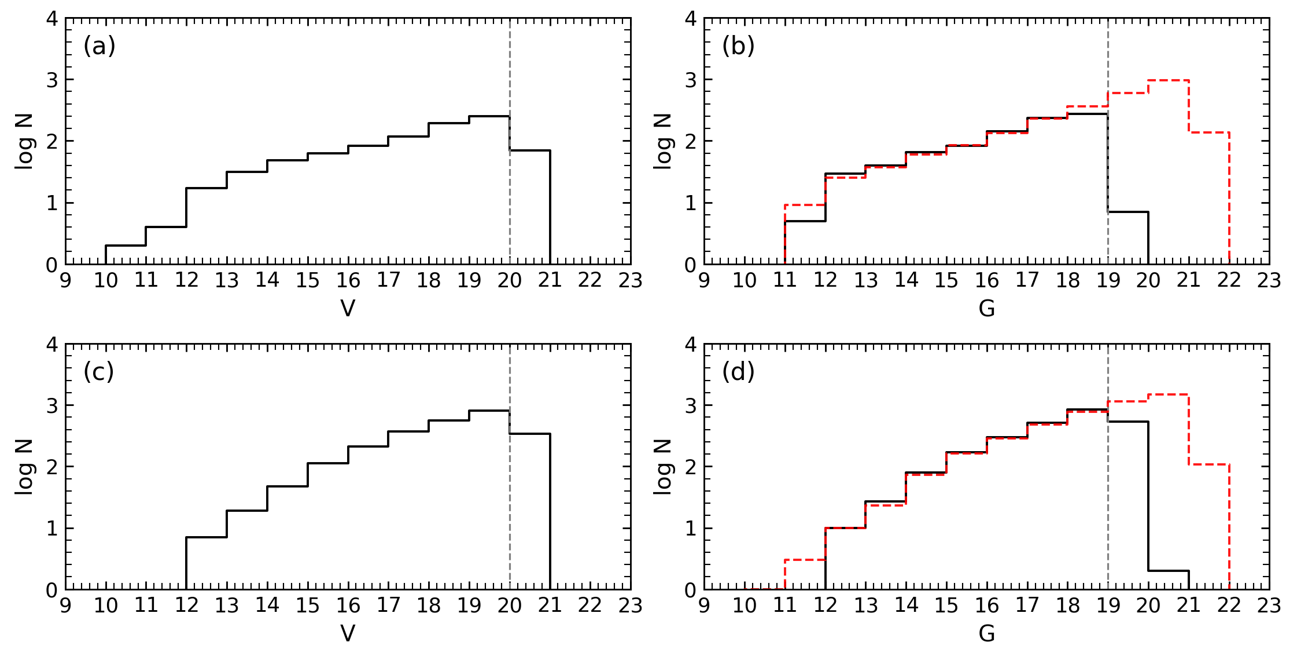}\\
\caption{Stellar counts of King 6 (a, b) and NGC 1605 (c, d) for per magnitude bin in $V$ and $G$ bands. The vertical grey dashed lines show the adopted faint limiting apparent magnitudes in $V$ and $G$ bands. Black lines represent the star counts in $V$ magnitude, whereas red dashed lines are counts in {\it Gaia} DR3 data for the same cluster regions.} 
\label{fig:histograms}
\end {figure*}

In this study, the photometric completeness limit is needed as part of the information to derive cluster parameters such as luminosity and mass functions, stellar density distributions, etc. Histograms of stellar counts, as functions by magnitude bin in $V$ and $G$, were used to determine the photometric completeness limits for the studied clusters. We compared the number of stars found with those counts gathered from {\it Gaia} DR3 data with counts of the same regions in the ground-based photometry. The {\it Gaia} DR3 data were prepared by considering equatorial coordinates given by \citet{Cantat-Gaudin20} and stellar magnitudes to the range $8<G<22$ mag. The stellar counts in $V$ and $G$ magnitude bins are shown in Fig. ~\ref{fig:histograms}: the black solid lines are the observational stellar distributions per $V$ and $G$ magnitude bins, while the red dashed lines (see in Figs.~\ref{fig:histograms}b and \ref{fig:histograms}d) indicate the {\it Gaia} DR3-based stellar counts. Vertical dashed lines represent the completeness ``turn-over'' magnitudes where the number of stars detected begins to drop (with increasing magnitude), indicating where completeness starts to affect the stellar counts. It can be seen from Figs. ~\ref{fig:histograms}b and \ref{fig:histograms}d that the number of stars detected in the study is in good agreement with the stellar distribution from {\it Gaia} DR3 data until the adopted completeness limits. These limits are $V=20$ mag for both King 6 and NGC 1605, which corresponds to $G=19$ mag. The number of stars of similar magnitude ranges that will be detected in a given image is related to the properties of CCD-telescope combinations and exposure times used in observations. Thus, the number of detected stars in fainter magnitudes will be different between ground and space-based observations, such as those used in the current study. This could contribute to the dissimilar stellar counts for magnitudes fainter than $G 
 = 19$ as seen in Figures~\ref{fig:histograms}b and \ref{fig:histograms}d.     

\begin{sidewaystable}
\setlength{\tabcolsep}{1.9pt}
\renewcommand{\arraystretch}{0.8}
\footnotesize
  \centering
 \caption{The catalogues for King 6 and NGC 1605. The complete table can be found electronically.}
    \begin{tabular}{cccccccccccc}
\hline
\multicolumn{11}{c}{King 6}\\
\hline
ID	 & RA           &	DEC	        &      $V$	    &	$U-B$      & $B-V$	      &	$G$	          & 	$G_{\rm BP}-G_{\rm RP}$	 & 	$\mu_{\alpha}\cos\delta$ & 	$\mu_{\delta}$ & 	$\varpi$	& $P$ \\

	 & (hh:mm:ss.ss)           &	(dd:mm:ss.ss)	&      (mag)	    &	(mag)      & (mag)	      &	(mag)	          & 	(mag)	 & 	(mas yr$^{-1}$) & 	(mas yr$^{-1}$) & 	(mas)	&  \\
\hline
0001 & 03:26:48.64 & +56:33:41.49 & 19.704(0.024) & ---          & 1.710(0.061) & 18.527(0.003) & 2.343(0.044) &  1.645(0.191) & -1.701(0.159) & 0.557(0.174) & 0.00 \\
0002 & 03:26:48.69 & +56:21:21.03 & 20.033(0.035) & ---          & 1.723(0.087) & 18.804(0.004) & 2.454(0.059) &  1.345(0.280) &  2.903(0.252) & 0.799(0.256) & 0.00 \\
0003 & 03:26:49.20 & +56:24:05.73 & 19.640(0.024) & ---          & 2.191(0.081) & 18.657(0.003) & 2.270(0.068) &  9.486(0.219) & -5.098(0.183) & 0.615(0.192) & 0.00 \\
0004 & 03:26:49.22 & +56:23:29.49 & 17.130(0.004) & 1.309(0.043) & 1.405(0.007) & 16.643(0.003) & 1.782(0.013) &  4.000(0.060) & -1.864(0.057) & 1.328(0.057) & 0.99 \\
...  & ...         & ...	      & ...           & ...          & ...          & ...           & ...          & ...           & ...           & ...          & ...  \\
...  & ...         & ...	      & ...           & ...          & ...          & ...           & ...          & ...           & ...           & ...          & ...  \\
...  & ...         & ...	      & ...           & ...          & ...          & ...           & ...          & ...           & ...           & ...          & ...  \\
0881 & 03:29:15.93 & +56:20:09.31 & 18.448(0.009) & ---          & 1.770(0.022) & 17.453(0.003) & 2.224(0.022) &  4.567(0.103) &  0.895(0.098) & 0.698(0.096) & 0.00 \\
0882 & 03:29:17.03 & +56:19:59.99 & 18.017(0.006) & 0.709(0.067) & 1.551(0.014) & 17.167(0.003) & 2.036(0.018) &  4.761(0.084) & -4.904(0.079) & 0.650(0.074) & 0.00 \\
0883 & 03:29:17.22 & +56:33:48.74 & 18.082(0.007) & ---          & 1.725(0.016) & 17.187(0.003) & 2.227(0.017) & -2.319(0.088) &  1.364(0.080) & 0.918(0.098) & 0.00 \\
0884 & 03:29:17.23 & +56:22:22.93 & 14.735(0.001) & 0.559(0.005) & 1.246(0.002) & 14.247(0.005) & 1.740(0.011) &  0.452(0.114) & -0.939(0.117) & 0.764(0.110) & 0.00 \\
\hline
\multicolumn{11}{c}{NGC 1605}\\
\hline
ID	 & RA           &	DEC	        &      $V$	    &	$U-B$      & $B-V$	      &	$G$	          & 	$G_{\rm BP}-G_{\rm RP}$	 & 	$\mu_{\alpha}\cos\delta$ & 	$\mu_{\delta}$ & 	$\varpi$	& $P$ \\
	 & (hh:mm:ss.ss)           &	(dd:mm:ss.ss)	&      (mag)	    &	(mag)      & (mag)	      &	(mag)	          & 	(mag)	 & 	(mas yr$^{-1}$) & 	(mas yr$^{-1}$) & 	(mas)	&  \\
\hline
0001 & 04:33:36.05  & +45:20:59.99 & 17.737(0.008)  & 1.265(0.086) & 1.283(0.012)  & 17.202(0.003) & 1.650(0.011) & 2.534(0.088)  &  0.951(0.062) &  0.881(0.080)  & 0.00 \\
0002 & 04:33:36.09  & +45:17:03.77 & 18.017(0.009)  & 0.975(0.086) & 1.285(0.014)  & 17.487(0.003) & 1.705(0.016) & 0.260(0.113)  &  0.394(0.079) &  0.556(0.100)  & 0.00 \\
0003 & 04:33:36.11  & +45:06:56.59 & 20.279(0.057)  & ---          & 1.260(0.087)  & 19.435(0.004) & 1.816(0.069) & 0.023(0.421)  &  0.047(0.306) & -0.087(0.419)  & 0.00 \\
0004 & 04:33:36.12  & +45:13:31.45 & 18.289(0.011)  & 0.658(0.089) & 1.372(0.018)  & 17.712(0.003) & 1.791(0.015) & 1.813(0.128)  & -1.426(0.090) &  0.322(0.110)  & 0.00 \\
...  & ...          & ...	       & ...            & ...          & ...           & ...           & ...          & ...           & ...           &  ...           & ...  \\
...  & ...          & ...	       & ...            & ...          & ...           & ...           & ...          & ...           & ...           &  ...           & ...  \\
...  & ...          & ...	       & ...            & ...          & ...           & ...           & ...          & ...           & ...           &  ...           & ...  \\
2471 & 04:35:33.05  & +45:15:23.51 & 19.805(0.038)  & ---          & 1.859(0.078)  & 18.837(0.003) & 2.095(0.063) & 6.042(0.299)  & -5.366(0.215) &  0.238(0.241)  & 0.00 \\
2472 & 04:35:33.08  & +45:11:27.28 & 18.765(0.016)  & ---          & 1.479(0.026)  & 18.105(0.003) & 1.954(0.028) & 1.475(0.255)  &  2.852(0.161) &  0.773(0.182)  & 0.00 \\
2473 & 04:35:33.13  & +45:17:50.24 & 17.714(0.007)  & 0.915(0.074) & 1.540(0.013)  & 16.987(0.003) & 2.011(0.013) & 1.526(0.082)  &  2.385(0.070) &  0.313(0.071)  & 0.00 \\
2474 & 04:35:33.37  & +45:21:49.32 & 20.000(0.048)  & ---          & 1.448(0.079)  & 18.998(0.005) & 2.001(0.057) & 0.875(0.346)  & -0.459(0.258) &  0.691(0.284)  & 0.00 \\
\hline
    \end{tabular}
      \label{tab:all_cat}%
\end{sidewaystable} 

\begin{table*}
\setlength{\tabcolsep}{3pt}
\renewcommand{\arraystretch}{0.8}
  \centering
  \caption{Mean internal photometric errors per magnitude bin in $V$ brightness.}
    \begin{tabular}{ccccccc|ccccccc}
      \hline
    \multicolumn{7}{c}{King 6} & \multicolumn{6}{c}{NGC 1605} \\
    \hline
  $V$ & $N$ & $\sigma_{\rm V}$ & $\sigma_{\rm U-B}$ & $\sigma_{\rm B-V}$ & $\sigma_{\rm G}$ &  $\sigma_{G_{\rm BP}-G_{\rm RP}}$ & $V$ & $N$ & $\sigma_{\rm V}$ & $\sigma_{\rm U-B}$ & $\sigma_{\rm B-V}$ & $\sigma_{\rm G}$ & $\sigma_{G_{\rm BP}-G_{\rm RP}}$\\
  \hline
  (08, 12] &   7 & 0.001 & 0.001 & 0.001 & 0.003 & 0.007 & (08, 12] & --- & ---   &  ---  & ---   & ---   & ---  \\
  (12, 14] &  48 & 0.002 & 0.003 & 0.003 & 0.003 & 0.005 & (12, 14] &  26 & 0.001 & 0.003 & 0.001 & 0.003 & 0.006\\
  (14, 15] &  48 & 0.001 & 0.005 & 0.002 & 0.003 & 0.005 & (14, 15] &  47 & 0.001 & 0.006 & 0.002 & 0.003 & 0.005\\
  (15, 16] &  65 & 0.001 & 0.011 & 0.003 & 0.003 & 0.006 & (15, 16] & 114 & 0.002 & 0.012 & 0.003 & 0.003 & 0.006\\
  (16, 17] &  85 & 0.002 & 0.025 & 0.006 & 0.003 & 0.008 & (16, 17] & 213 & 0.003 & 0.025 & 0.005 & 0.003 & 0.007\\
  (17, 18] & 117 & 0.004 & 0.054 & 0.012 & 0.003 & 0.012 & (17, 18] & 373 & 0.006 & 0.050 & 0.010 & 0.003 & 0.011\\
  (18, 19] & 196 & 0.009 & 0.074 & 0.025 & 0.003 & 0.025 & (18, 19] & 574 & 0.013 & 0.106 & 0.021 & 0.003 & 0.023\\
  (19, 20] & 254 & 0.021 & 0.087 & 0.066 & 0.003 & 0.054 & (19, 20] & 818 & 0.029 & 0.173 & 0.051 & 0.004 & 0.048\\
  (20, 22] &  64 & 0.033 & 0.156 & 0.107 & 0.004 & 0.077 & (20, 22] & 309 & 0.052 & 0.239 & 0.101 & 0.004 & 0.073\\
      \hline
    \end{tabular}%
  \label{tab:phiotometric_errors}%
\end{table*}%

\subsection{Structural Parameters of the Clusters}

\begin{figure}[t]
\centering
\includegraphics[scale=0.65, angle=0]{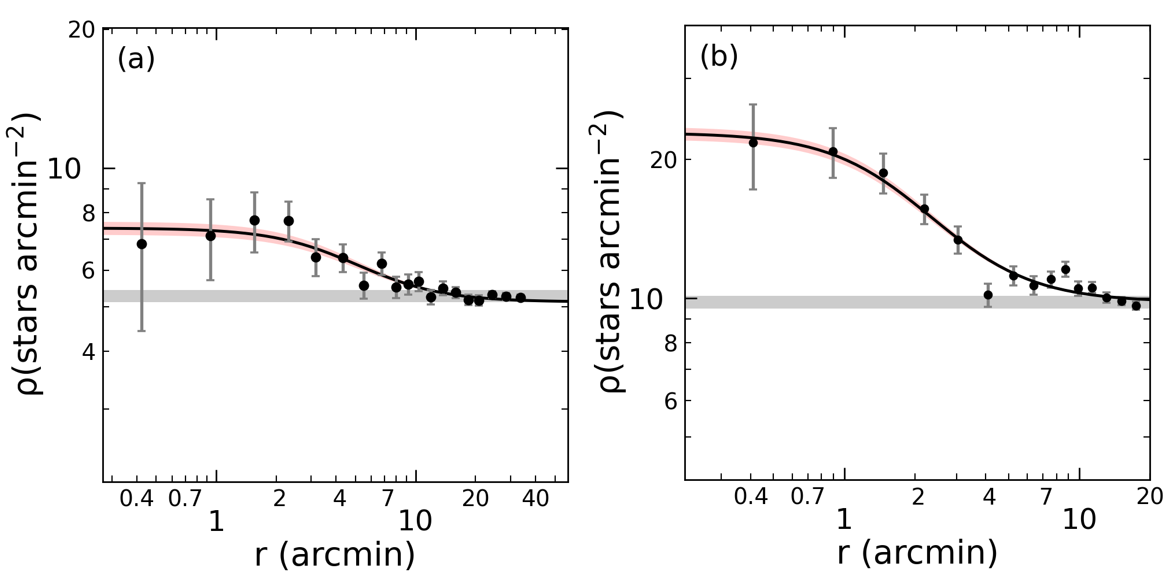}\\
\caption{The radial density profiles of King 6 (a) and NGC 1605 (b) OCs. The fitted black curve in each diagram is the \citet{King62} profile, whereas the horizontal grey band represents the background stellar density. The red-shaded domain shows the $1\sigma$ King fit uncertainty.} 
\label{fig:king}
\end {figure}

The structural parameters for King 6 and NGC 1605 were obtained through radial density profile (RDP) analyses. To construct an RDP, we used the {\it Gaia} DR3 data set due to its unlimited field of view. For each cluster, we retrieved the sources within a radius of $40'$ centered at the coordinates given by \citet{Cantat-Gaudin20}. Considering only the stars brighter than the photometric completeness limits ($G=19$ mag for both clusters), we counted stars in a series of concentric rings centered on the adopted cluster centers \citep{Cantat-Gaudin20} and so derived stellar densities ($\rho$). These values were calculated by dividing stellar counts in each ring by the areas of the appropriate ring. In order to parameterize the stellar densities, the RDP of \citet{King62} was fitted to the calculated stellar densities using a $\chi^2$ minimization procedure. The \citet{King62} profile is expressed by the equation $\rho(r)=f_{\rm bg}+[f_{\rm 0}/(1+(r/r_{\rm c})^2)]$, where $f_{\rm bg}$, $f_{\rm 0}$, $r_{\rm c}$ and $r$ indicate background density, the central density, core radius, and the radius from the cluster centre, respectively. The best-fitting RDP models of stellar density distribution versus radius from the cluster centre are plotted in Fig.~\ref{fig:king}. We calculated the central stellar density and background stellar density of the clusters, together with the core radius, as:
\begin{itemize}
    \item{$f_{\rm 0}=2.28\pm 0.24$ and $f_{\rm bg}=5.12\pm 0.16$ stars arcmin$^{-2}$, and $r_{\rm c}=4.68\pm 1.07$ arcmin for King 6, and}
    \item{$f_{\rm 0}=13.05\pm 0.73$ and $f_{\rm bg}=9.81\pm 0.30$ stars arcmin$^{-2}$, and $r_{\rm c}=1.90\pm 0.20$ arcmin for NGC 1605.}
\end{itemize}
In Fig.~\ref{fig:king}, the best-fitting RDP is shown by a black solid line, while the horizontal grey band is the level of the background density. We obtained the limiting radius ($r_{\rm lim}^{\rm obs}$) for each cluster through visual inspection,  considering the RDP model fit and background density. The stellar density is above the background levels up to 10 arcmin for both clusters (see Fig.~\ref{fig:king}). Hence we adopted the limiting radii as $r_{\rm lim}^{\rm obs}=10'$ for King 6 and NGC 1605. We used only the stars inside these limiting radii for further analyses. To compare the accuracy of the observed limiting radii, we also estimated limiting radii ($r_{\rm lim}^{\rm cal}$) using the mathematical expression given by \citet{Bukowiecki11}:
\begin{equation}
r_{\rm lim}^{\rm cal}=r_{\rm c}\sqrt{\frac{f_0}{3\sigma_{\rm bg}}-1}
\end{equation}
The calculated limiting radii ($r_{\rm lim}^{\rm cal}$) were found to be $9\,'\!.1$ and $9\,'\!.3$ for King 6 and NGC 1605, respectively. These values are in good agreement with the observational values. 

\subsection{Member Selection and Color-Magnitude Diagrams}
\label{section:cmds}

Since OCs are located along the Galactic plane, they are affected by field star contamination. This complicates the confident detection and identification of physical members (of the OC), which in turn influences the determination of reliable astrophysical parameters of the open clusters. Therefore the effect of field stars should be identified and eliminated. Cluster member stars have similar vectoral movements as a result of being formed from the same molecular cloud. Using this property offers a path to identify cluster physical members \citep{Yadav13,Bisht20}, as explained in this section. In the study, we used equatorial coordinates, proper-motion components, and trigonometric parallaxes of the stars to derive cluster members. Such an astrometric approach to determine cluster membership has been successfully applied by various researchers \citep{Vasilevskis58, Stetson80, Zhao90, Balaguer98, Sarro14, Krone-Martins14, Olivares19, Pera21}.  

We used the Unsupervised Photometric Membership Assignment in Stellar Clusters \citep[{\sc UPMASK};][]{Krone-Martins14} algorithm, as implemented in the {\sc pyUPMASK} \citep{Pera21} package, to identify members of the King 6 and NGC 1605 OCs. {\sc pyUPMASK} is written in Python and has a general structure that follows the algorithm of {\sc UPMASK} \citep{Pera21}. Both algorithms were previously used to calculate membership probabilities of open clusters by various researchers \citep[cf.][]{Koc22, Tasdemir23, Yontan23c}. 

We applied {\sc pyUPMASK} to calculate the membership probabilities ($P$) of stars detected in the regions of King 6 and NGC 1605. We ran the program in a five-dimensional parametric space which contains equatorial coordinates ($\alpha$, $\delta$), proper-motion components ($\mu_{\alpha}\cos \delta$, $\mu_{\delta}$), trigonometric parallaxes ($\varpi$) and relevant uncertainties of detected stars. By running 1000 iterations we found 148 (for King 6) and 240 (for NGC 1605) likely cluster members. These stars have cluster membership probabilities equal to or greater than 50\% (which we considered to be the lower limit for membership probability).

\begin{figure*}[t]
\centering
\includegraphics[scale=0.76, angle=0]{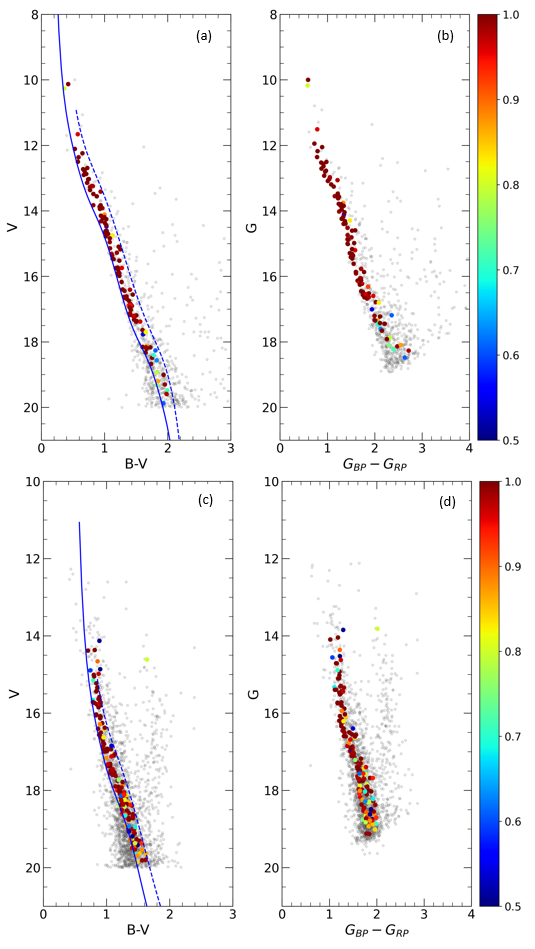}
\caption{Color-magnitude diagrams of King 6 (a, b) and NGC 1605 (c, d) based on {\it UBV} (a, c) and {\it Gaia} DR3 (b, d). The blue lines represent the ZAMS \citep{Sung13} including the MS broadening. The membership probabilities of stars are shown with different colors. These member stars lie within $r_{\rm lim}^{\rm obs}=10'$ of the cluster centres obtained for both clusters. The stars with low membership probabilities are plotted as grey dots. 
\label{fig:cmds}}
\end {figure*}

We therefore plotted $V\times (B-V)$ CMDs using the most likely cluster members ($P\geq0.5$) and fitted a ZAMS from \citet{Sung13} to the cluster sequence visually. The ZAMS was shifted by 0.75 mag towards brighter magnitudes to allow equal mass binary stars as members. We considered the $V$ magnitude limits and the cluster $r_{\rm lim}$ radii, along with the ZAMS fitting, to estimate 112 stars as `most likely' members ($P\geq0.5$) of King 6 and 160 for NGC 1605. We subsequently used these stars to determine astrophysical parameters for each cluster, as discussed below. The $V\times (B-V)$ and $G\times (G_{\rm BP}-G_{\rm RP})$ diagrams are given as Fig.~\ref{fig:cmds}. Figs.~\ref{fig:cmds}a (for King 6) and \ref{fig:cmds}c (for NGC 1605) are {\it UBV}-based CMDs. These show the distribution of field stars and the stars considered to be the most probable cluster members, together with the fitted ZAMS. Figs.~\ref{fig:cmds}b (for King 6) and \ref{fig:cmds}d (for NGC 1605) show the distribution of the most likely cluster members on {\it Gaia}-based photometry.

\begin{figure*}[!t]
\centering
\includegraphics[scale=0.9, angle=0]{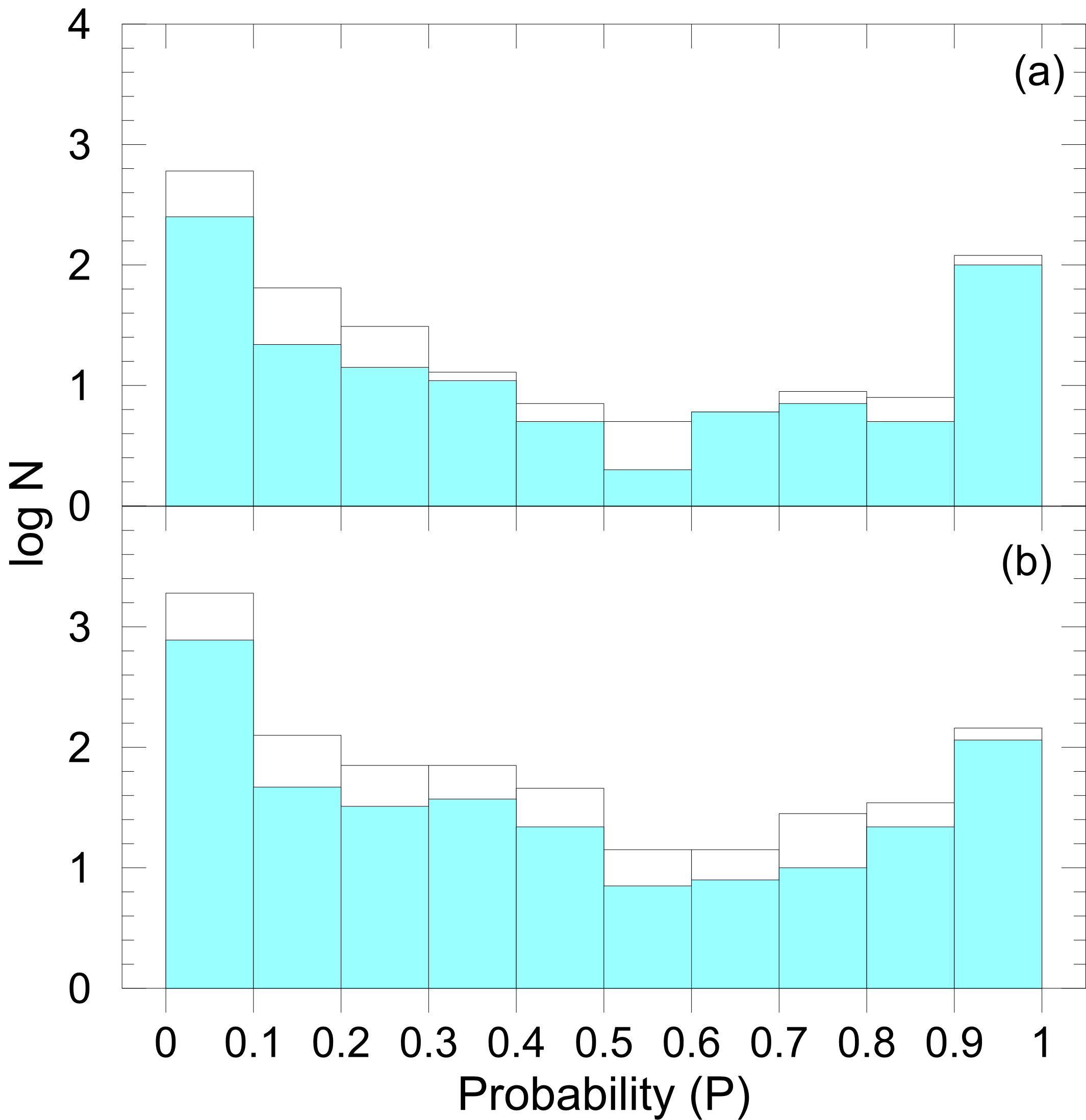}
\caption{Distribution of cluster membership probabilities for the stars in the direction of King 6 (a) and NGC 1605 (b). The white shaded bars are the counts of stars which have been detected in the cluster regions. The cyan-colored shaded bars represent the number of stars that lie within the main-sequence band and the cluster limiting radius $r_{\rm lim}^{\rm obs}$.
\label{fig:prob_hists} }
\end {figure*}

\begin{figure}[t]
\centering
\includegraphics[scale=.45, angle=0]{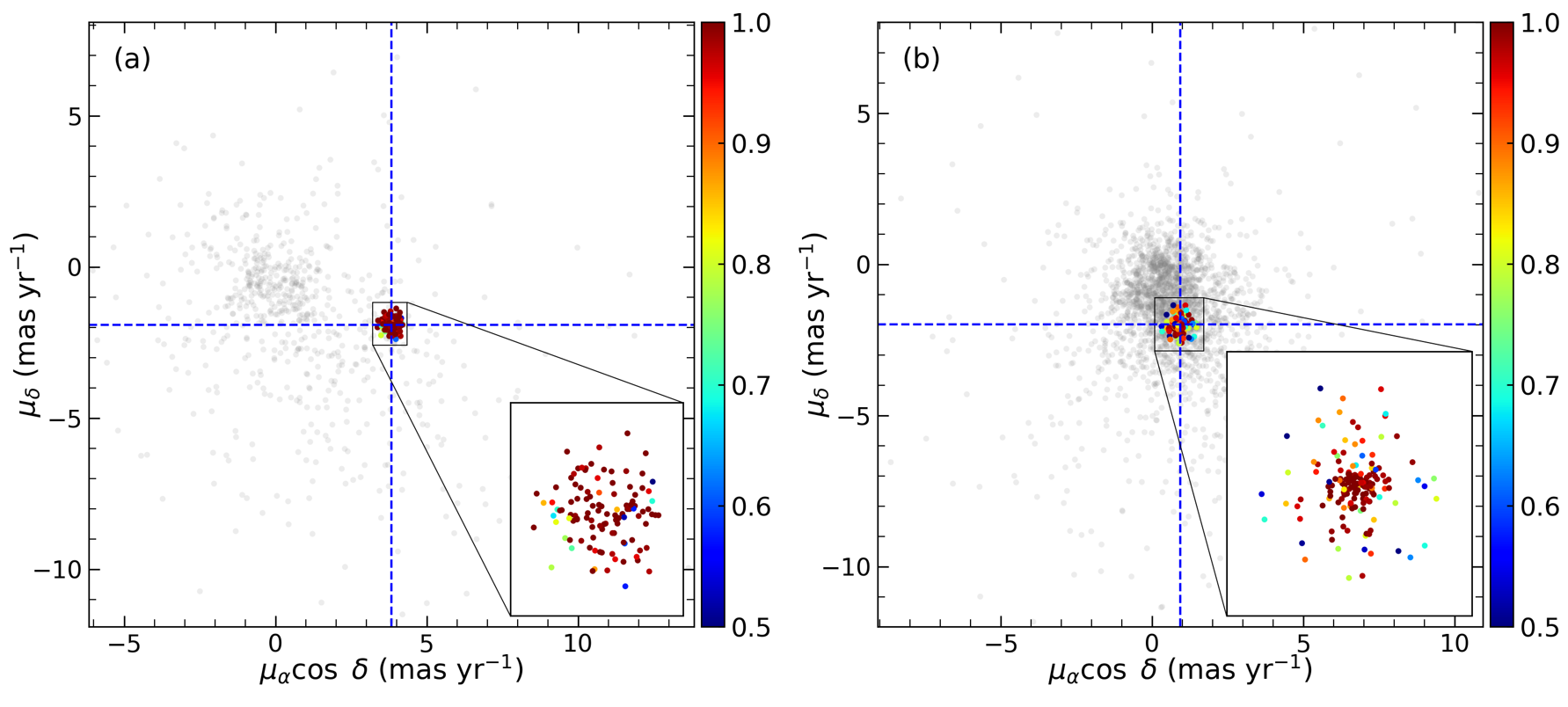}\\
\caption{{\it Gaia} DR3 astrometry-based vector point diagrams of King 6 (a) and NGC 1605 (b). The membership probabilities of the stars are represented with the color scale shown on the right side. The zoomed boxes in panels (a) and (b) indicate the region of concentration for each cluster in the diagrams. The intersection of the dashed blue lines indicates the mean proper motion values.
\label{fig:VPD_all}} 
 \end {figure}

\begin{figure}[t]
\centering
\includegraphics[scale=.55, angle=0]{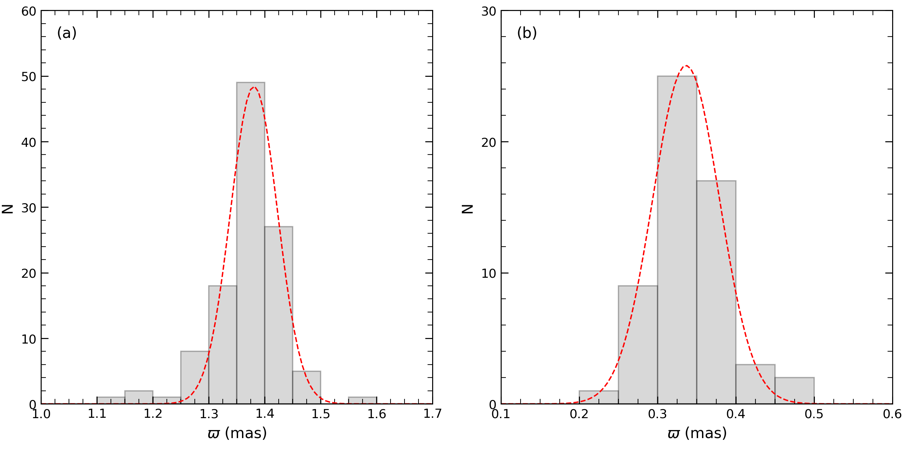}\\
\caption{{\it Gaia} DR3-based trigonometric parallax histograms for King 6 (a) and NGC 1605 (b). The red-dashed curve shows the Gaussian fit applied to the distributions.
\label{fig:plx_hist}}
\end {figure}

The membership probability distributions (for detected stars in the cluster regions) are shown in Fig.~\ref{fig:prob_hists}. We constructed vector-point diagrams (VPDs), as shown in Fig.~\ref{fig:VPD_all}. It is clear from that figure that the most likely cluster members for the two clusters are associated with central points to the clusters. We can interpret that King 6 (Fig.~\ref{fig:VPD_all}a) is clearly more separated from the scattered field stars than NGC 1605 is (Fig.~\ref{fig:VPD_all}b). Considering only the most likely member stars, we obtained values for mean proper-motion components. These are shown as the intersections of the blue dashed lines in Fig.~\ref{fig:VPD_all}. For King 6 we calculated the mean proper motion as ($\mu_{\alpha}\cos\delta$, $\mu_{\delta}$)=($3.833\pm0.034$, $-1.906\pm0.032$) mas yr$^{-1}$ and for NGC 1605 as ($\mu_{\alpha}\cos\delta$, $\mu_{\delta}$)=($0.928\pm0.104$, $-1.997\pm0.082$) mas yr$^{-1}$. These findings are in good agreement with the values produced with {\it Gaia} data for both of the clusters (see Table~\ref{tab:literature}). We also estimated mean trigonometric parallaxes $\varpi$ by fitting Gaussians to the histograms of trigonometric parallax (Fig.~\ref{fig:plx_hist}). During these analyses, we used the stars having $P\geq 0.5$ membership probabilities and precise parallax accuracy ($\sigma_{\varpi}/\varpi < 0.2$). This resulted in estimates that the mean $\varpi$ is $1.381\pm 0.042$ mas and $0.336\pm 0.043$ mas for King 6 and NGC 1605, respectively. By applying the linear equation $d({\rm pc})=1000/\varpi$ (mas) to these mean values, we derived parallax distances as $d_{\varpi}=724\pm 22$ pc for King 6 and $d_{\varpi}=2976\pm 381$ pc for NGC 1605. Fig.~\ref{fig:plx_hist} shows the distribution for trigonometric parallaxes of member stars with a Gaussian fit (red dashed lines) applied to each data set.  

\newpage
\section{Basic Parameters of the Open Clusters}
This section summarizes the procedures for the astrophysical analyses of King 6 and NGC 1605. We used two-color diagrams (TCDs) to calculate the reddening and photometric metallicities separately.  Keeping these two parameters as constants and using CMDs, we next obtained the distance moduli and ages simultaneously \citep[as performed in previous studies cf.][]{Bostanci15,  Bostanci18}. 

\subsection{Color Excess through the Cluster Region}
\label{sec:reddening}
\begin{figure*}
\centering
\includegraphics[scale=0.5, angle=0]{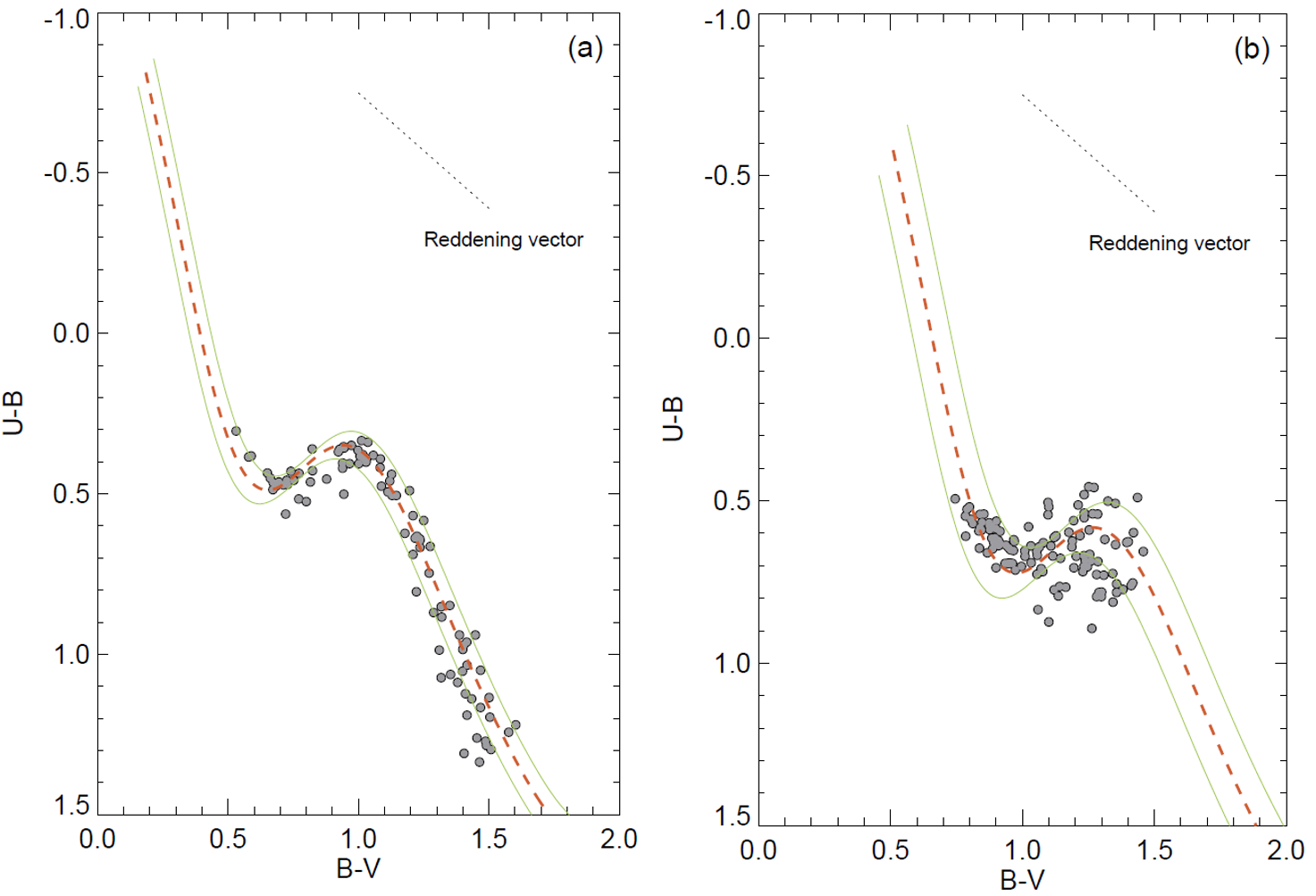}
\caption{The $(U-B)\times(B-V)$ diagrams of King 6 (a) and NGC 1605 (b). The red dashed lines plot the reddened ZAMS given by \citet{Sung13}. The green solid lines indicate the $\pm1\sigma$ standard deviations range limits. The reddening vector is represented by the dashed grey line. 
\label{fig:tcds}} 
\end{figure*}

The color excesses of the two open clusters are estimated by plotting $(U-B)\times (B-V)$ TCDs. We considered the stars located within limiting radii ($r_{\rm lim}^{\rm obs}\leq10'$) and with membership probabilities greater than 0.5. Stars in the clusters' main sequences were selected for the color excess analyses. With these limitations, we have selected the most likely main-sequence stars within $12\leq V \leq 17$ mag for King 6 and $14.75\leq V \leq 20$ mag for NGC 1605. The selected stars were compared with the de-reddened ZAMS of solar-metallicity \citep{Sung13} on $(U-B)\times (B-V)$ TCDs (Fig.~\ref{fig:tcds}). The ZAMS was shifted along the slope of the reddening vector $\alpha=E(U-B)/E(B-V)=0.72$ presented by \citet{Johnson53} through the use of $\chi^2$ minimization. Hence, the best-fit solutions indicated that the color excess $E(B-V)$ is $0.515\pm 0.030$ mag for King 6 and $0.840\pm 0.054$ mag for NGC 1605. Fig.~\ref{fig:tcds} shows the best fitting ZAMS (red-dashed lines) for the two clusters.


\begin{figure*}
\centering
\includegraphics[scale=0.6, angle=0]{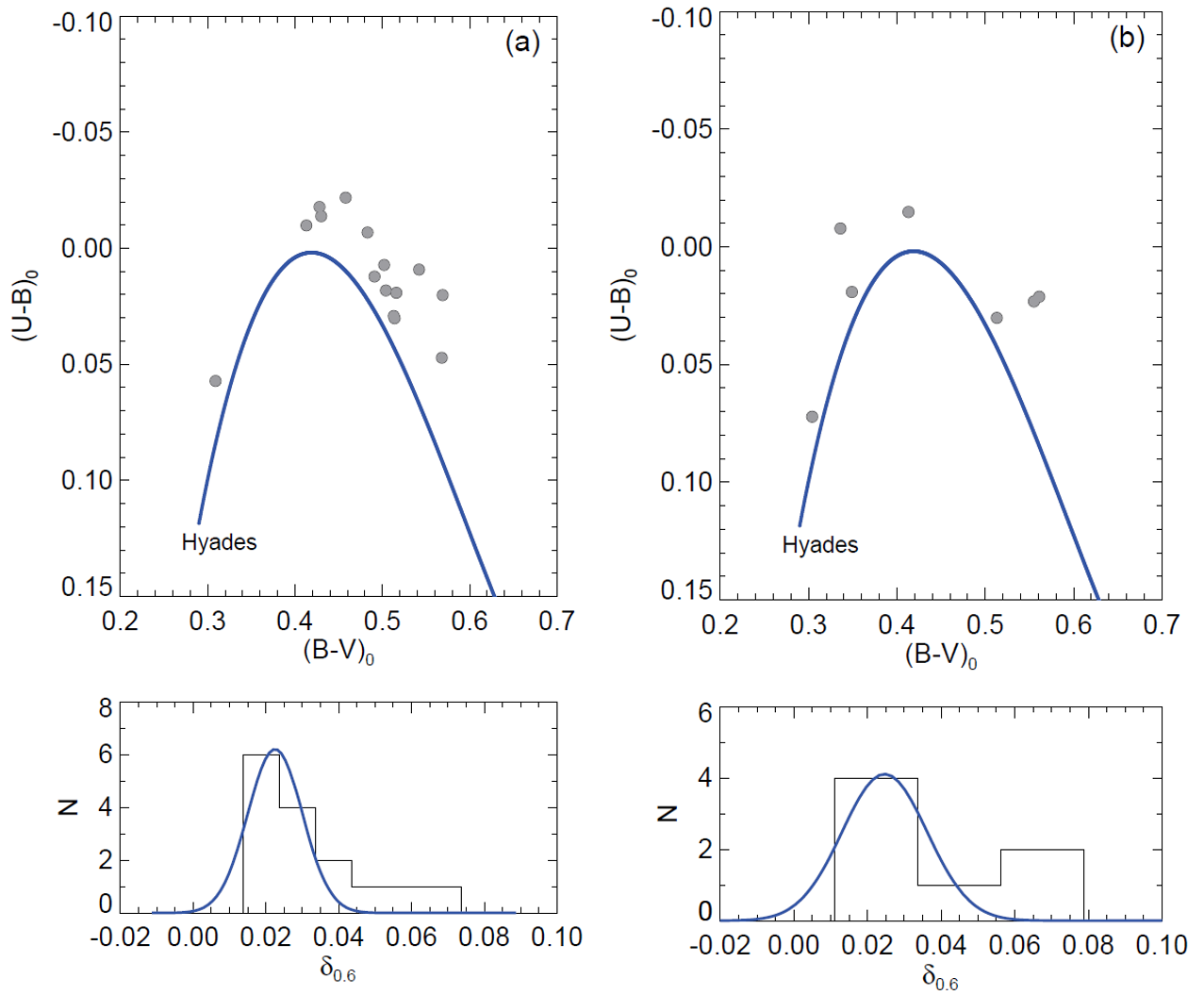}
\caption{The $(U-B)_0\times(B-V)_0$ TCDs (upper panels) and histograms for the normalized $\delta_{0.6}$ (lower panels) for 15 (King 6, panel a) and 7 (NGC 1605, panel b)  F-G type main-sequence stars statistically consider most probably cluster members. The solid blue lines in the TCDs represent the main sequence of Hyades(in the upper subfigures) and Gaussian fits (in the lower subfigures).
\label{fig:hyades}} 
\end {figure*}

\subsection{Photometric Metallicity}
\label{sec:metallicity}

To determine the [Fe/H] photometric metallicities for King 6 and NGC 1605, we adopted the $U\!BV$ data-based method of \citet{Karaali11}. This method uses the UV-excesses of F and G spectral type main-sequence stars, which is consistent with the $0.3\leq (B-V)_0\leq0.6$ color range. To apply this method, initially we estimated de-reddened $(B-V)_0$ and $(U-B)_0$ values of the most likely cluster main-sequence stars ($P\geq 0.5$) by using $E(B-V)$ and $E(U-B)$ color excesses derived in the study, then we limited the calculated $(B-V)_0$ color index within $0.3\leq (B-V)_0\leq0.6$ mag \citep{Eker18, Eker20} to select F and G type main-sequence stars. Hence, we reached 15 and 7 stars with membership probabilities greater than 0.5 for King 6 and NGC 1605, respectively. We constructed $(U-B)_0\times(B-V)_0$ diagrams for these stars, along with the Hyades main-sequence stars, to compare their $(U-B)_0$ values corresponding to the same de-reddened $(U-B)_0$ data. The upper panel of Fig.~\ref{fig:hyades},  $(U-B)\times(B-V)$ TCDs of the selected stars and Hyades main sequence. The comparison is described as UV-excess ($\delta$), given by the expression $\delta =(U-B)_{\rm 0,H}-(U-B)_{\rm 0,S}$ where the H and S subscripts indicate Hyades and cluster stars, respectively. The photometric metallicity calibration is defined within the $0.3\leq (B-V)_0\leq0.6$ color range and reaches its maximum UV-excess at $(B-V)=0.6$ mag. Therefore, the UV-excess' calculated for F-G type main-sequence stars needs to be normalized by a factor $f$, which is defined as the guillotine factor \citep{Sandage69}. In order to employ \citet{Karaali11}'s method we estimated the normalised UV-excess of the modelled stars to the UV-excess at $(B-V)_0 = 0.6$ mag (i.e. $\delta_{0.6}$). By constructing histograms of normalized UV-excesses we fitted Gaussian functions to the distributions and obtained mean $\delta_{0.6}$ values for each cluster  \citep{Karaali03a, Karaali03b}. The fitted Gaussian functions to the normalised UV-excesses for each cluster are shown in the lower panels of Fig.~\ref{fig:hyades}. The result of the fitting procedure provides the mean $\delta_{0.6}$ value as $\delta_{0.6}=0.023\pm0.007$ mag for King 6 and $\delta_{0.6}=0.025\pm0.012$ for NGC 1605. The $\pm1\sigma$ standard deviation of the Gaussian fit gives the uncertainty of the mean $\delta_{0.6}$. We estimated the photometric metallicity of each cluster by considering the mean $\delta_{0.6}$ value in the expression of \citet{Karaali11}:   
\begin{eqnarray}
{\rm [Fe/H]}=-14.316(1.919)\delta_{0.6}^2-3.557(0.285)\delta_{0.6}+0.105(0.039).
\end{eqnarray}

Taking into account the internal errors of the photometric metallicity calibration, as well as the photometric errors of the cluster member stars and the uncertainties in the cluster's color excesses, the external errors in the metallicity have an uncertainty of about 0.19 dex. Internal errors due to calibration ($\pm 0.06$ dex) and errors due to photometric measurements ($\pm 0.19$ dex) were evaluated together. Thus, we derived the photometric metallicities as [Fe/H]=$0.02\pm 0.20$ dex for King 6 and [Fe/H]=$0.01\pm 0.20$ dex for NGC 1605.  

In order to reliably select the isochrones that will be used for the determination of a cluster's age, one should transform the estimated [Fe/H] values to the mass fraction $z$. To do this, we utilized the following equations which are recommended for the {\sc parsec} isochrones of \citep{Bressan12} by Bovy\footnote{https://github.com/jobovy/isodist/blob/master/isodist/Isochrone.py}.  These are given as follows:
\begin{equation}
z_{\rm x}={10^{{\rm [Fe/H]}+\log \left(\frac{z_{\odot}}{1-0.248-2.78\times z_{\odot}}\right)}}
\end{equation}      
and
\begin{equation}
z=\frac{(z_{\rm x}-0.2485\times z_{\rm x})}{(2.78\times z_{\rm x}+1)}.
\end{equation} 
Here $z_{\rm x}$ and $z_{\odot}$ represent the intermediate and solar fraction values, respectively. The solar metallicity $z_{\odot}$ was adopted as 0.0152 \citep{Bressan12}. We obtained $z=0.016$ for King 6 and $z=0.015$ for NGC 1605.

\subsection{Distance Moduli and Age of the Clusters}
\label{sec:distance_age}

In this study, we plotted three CMDs combining $U\!BV$ and {\it Gaia} photometry to obtain the age and distance of each cluster. These results were also tested in terms of compatibility of the color excesses and metallicity values derived from TCDs.

The age and distance modulus were estimated together via fitting of {\sc parsec} isochrones by \citet{Bressan12}.  For {\sc parsec} isochrones, data from the {\it UBVRIJHK} \citep{Bessell90, Maiz06} and {\it Gaia} EDR3 \citep{{Riello21}} photometric systems in the CMD\footnote{CMD is a set of routines that provide interpolated isochrones in a grid, together with derivatives such as luminosity functions, simulated star clusters, etc. CMD 3.7 database was used in this study.}. The fitting process was done by visual inspection, taking into account the position of the most likely member stars ($P\geq 0.5$) on the CMDs. We selected isochrones of different ages that scaled to the metal fraction $z$ estimated for each cluster (Section~\ref{sec:metallicity}) and superimposed onto the $V\times (U-B)$, $V\times (B-V)$, and $G\times (G_{\rm BP}-G_{\rm RP})$ diagrams. We fitted the isochrone to the $V\times (U-B)$, $V\times (B-V)$ diagrams according to $E(U-B)$ and $E(B-V)$ values calculated in Section~\ref{sec:reddening}, whereas for the $G\times (G_{\rm BP}-G_{\rm RP})$ we fitted isochrones by using the selective absorption coefficients ($A_{\lambda}/A_{\rm V}$) for {\it Gaia} DR3 $G$, $G_{\rm BP}$, and $G_{\rm RP}$ pass-bands which were taken from \citet{Cardelli89} and \citet{Odennell94} as 0.83627, 1.08337 and 0.63439, respectively.   

\begin{figure*}
\centering
\includegraphics[scale=0.80, angle=0]{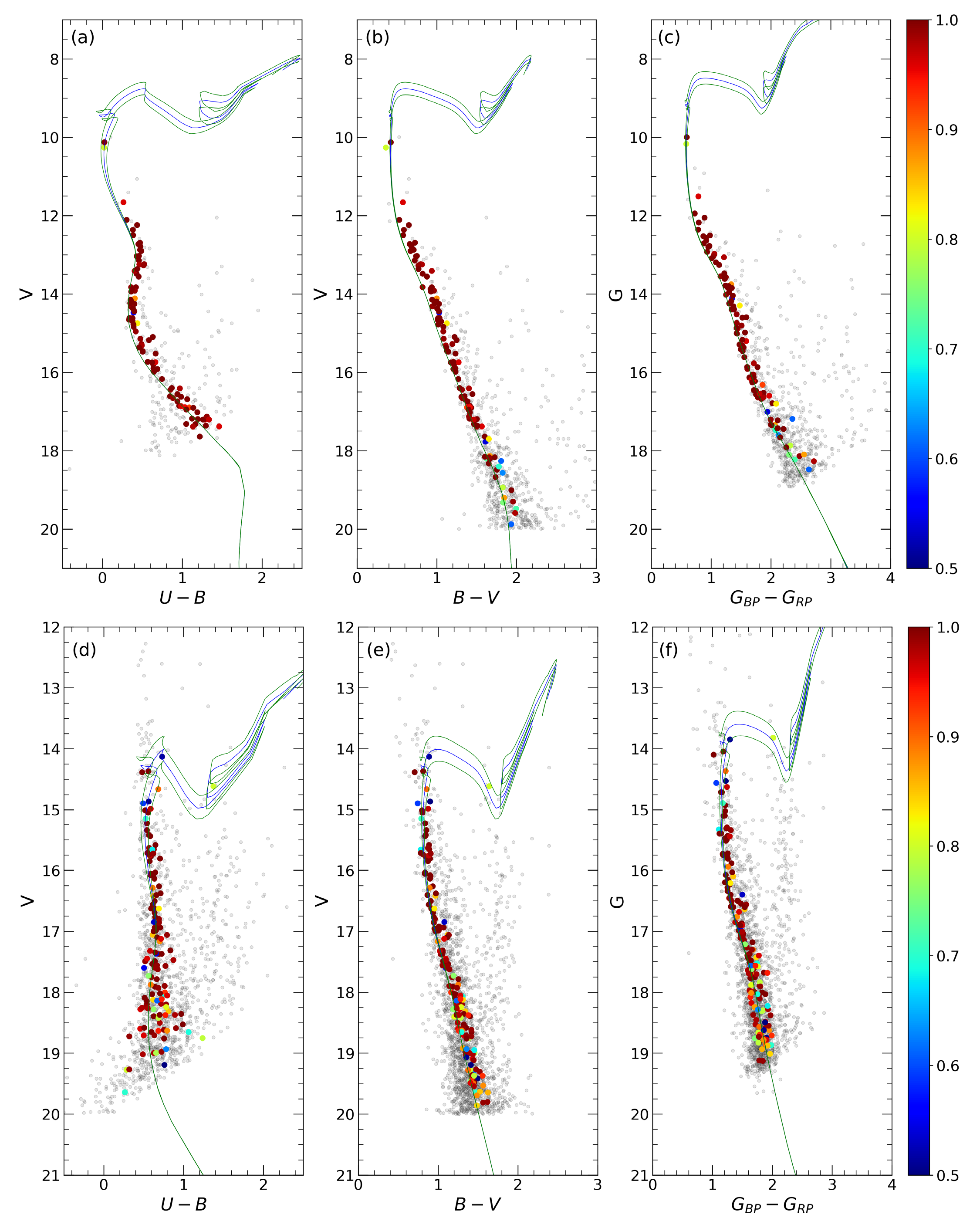}
\caption{$U\!BV$ and {\it Gaia} based CMDs for the King 6 (panels a, b, and c) and NGC 1605 (panels d, e, and f). Membership probabilities of the most probable cluster stars are represented with different color scales that are shown on the colorbars to the right, whereas field stars are shown as grey-colored dots. The best fitting {\sc parsec} isochrones and their errors are presented as the blue and green lines, respectively. Superimposed isochrone ages match to 200 Myr for King 6 and 400 Myr for NGC 1605.
\label{fig:figure_ten} }
\end {figure*}

The isochrone fitting procedure made reference to the positions of the most likely main sequence, turn-off point and giant members ($P\geq 0.5$) of each cluster. We considered the relation of \citet{Carraro17} for error estimates of distance moduli and distances. For the age uncertainties, we used low and high-age isochrones that well fit the observed scatter about the main sequence and turn-off. We present $V\times (U-B)$, $V\times (B-V)$ and $G\times (G_{\rm BP}-G_{\rm RP})$ diagrams with best fitting isochrones in Fig.~\ref{fig:figure_ten}. It can be seen from the Figs.~\ref{fig:figure_ten}-a, -b, and -c that the main sequence of King 6 shows about a 2 mag gap among the brightest stars. The fact that similar age values are given in the literature for King 6 shows that the isochrones well represent the cluster morphology. Therefore, the $\sim 2$ mag gap in main-sequence could be due to the lack of massive stars in the initial mass of the cluster. 

The following results were estimated from the isochrone fitting to the CMDs:
\begin{itemize}
\item{{\bf King 6}: By superimposing isochrones of $\log({\rm age}) = 8.26, 8.30$ and 8.34 with $z =0.016$ to the $U\!BV$ and {\it Gaia}-based CMDs, we obtained the apparent distance modulus as $\mu_{V}=10.892 \pm 0.099$ mag. This corresponds to the isochrone-based distance being $d_{\rm iso}=723\pm 34$ pc. We applied over-weights to the main-sequence and turn-off member stars, determining the cluster age to be $t=200\pm 20$ Myr. The estimated distance value matches well with most of the results that were obtained from {\it Gaia} data by various researchers (see Table~\ref{tab:literature} for a detailed comparison). The isochrone distance of King~6 is in good agreement with the {\it Gaia} DR3 trigonometric parallax distance value $d_{\varpi}=724\pm 22$ pc as estimated in Section~\ref{section:cmds}. The derived cluster age is consistent with the findings of \citet{Ann02}, \citet{Cantat-Gaudin20} and \citet{Dias21} (see Table~\ref{tab:literature} for detailed comparison).} 
\item{{\bf NGC 1605}: The same method was employed for this cluster, using isochrones for $\log({\rm age}) = 8.55, 8.60$, and 8.65 with $z = 0.015$. The distance modulus is $\mu_{\rm V}=15.028\pm 0.167$ mag, matching the isochrone-based distance of  $d_{\rm iso}=3054\pm 243$ pc. The cluster age is $t=400\pm 50$ Myr. Because of its age, during the calculation of the ages and distance moduli, we made reference to the turn-off and giant members. The estimated distance is compatible with the majority of results presented by various researchers listed in Table~\ref{tab:literature}. The distance value is within the errors of the trigonometric parallax distance value $d_{\varpi}=2976\pm 381$ pc (Section~\ref{section:cmds}). The estimated age of the cluster is in good agreement with the results of \citet{Loktin17} and \citet{Dias21}.} 
\end{itemize}
\noindent

\section{Kinematics and Galactic Orbit Integration}

The Python-based galactic dynamics library \citep[{\sc galpy}\footnote{See also https://galpy.readthedocs.io/en/v1.5.0/},][]{Bovy15} was used to determine the orbital properties of the two clusters according to the {\sc MWPotential2014} Galactic potential from \citet{Bovy15}. {\sc MWPotential2014} includes a module of an axisymmetric potential for the Milky Way galaxy and uses three components, which are bulge, disc, and halo potentials. The adopted bulge potential took the form from \citet{Bovy15} who presented it based on a spherical power-law density, while the disk is in the form of \citet{Miyamoto75} who defined an axisymmetric disk, and the halo is in the form stated by \citet{Navarro96} who described a spherically symmetric distribution of dark matter in the halo. The explicit parameters of {\sc MWPotential2014} that were used for the analyses of orbital integration and model fit parameter constraints are given in \citet {Bovy15} in detail. From that study, we adopted the galactocentric distance and orbital velocity as $R_{\rm GC}=8$ kpc and $V_{\rm rot}=220$ km s$^{-1}$, respectively \citep{Bovy15, Bovy12}. The vertical distance of the Sun from the Galactic plane was adopted to be $27 \pm 4$ pc as presented by \citet{Chen00}.

For complete orbit integration, the mean radial velocity is needed. In the current study, the radial velocity values for the most likely member stars were taken from {\it Gaia} DR3. We considered the stars' existing radial velocity measurements for those stars with membership probabilities of $P\ge 0.8$. Hence, we obtained 23 stars for King 6 and one star for NGC 1605 to estimate their mean radial velocities. Considering the method of weighted averages (for equations see \citeauthor{Soubiran18}, \citeyear{Soubiran18}) for the King 6 member stars, we calculated the mean value as $V_{\rm R}=-23.40\pm 3.26$ km s$^{-1}$, whereas for NGC 1605 we adopted the value of $V_{\rm R}=-15.27\pm 1.35$ km s$^{-1}$ based on {\it Gaia} DR3 measurement. The result for King 6 is in good agreement with the literature studies, being within 1-3 km s$^{-1}$ of the values listed in Table~\ref{tab:literature}. The radial velocity value adopted in this study for NGC 1605 is compatible with the result of \citet{Zhong20}, who calculated the mean value $V_{\rm R}=-12.033\pm17.417$ km s$^{-1}$ from eight stars considering their LAMOST DR5 data \citep{Luo19}. However, the result obtained in this study differs from the $V_{\rm  R}=-1.15\pm0.12$ km s$^{-1}$ value of \citet{Soubiran18} and \citet{Tarricq21}, who adopted the radial velocity of one cluster member from {\it Gaia} DR2 data \citep{Gaia18}. The large errors in the radial velocity measurements given for NGC 1605 in the study of  \citet{Zhong20} may be due to a binary star effect or low spectral resolution. 

To estimate the orbital parameters for each cluster, we utilized {\sc MWPotential2014} with the input parameters of equatorial coordinates ($\alpha$, $\delta$) taken from \citet{Cantat-Gaudin20}, the mean proper-motion components ($\mu_{\alpha}\cos\delta$, $\mu_{\delta}$) derived in Section~\ref{section:cmds}, isochrone distance ($d_{\rm iso}$) from Section~\ref{sec:distance_age}, and the radial velocity ($V_{\rm r}$) calculated in the study (see also Table~\ref{tab:Final_table}). Integration of the orbits was carried out backward in time with 1 Myr steps up to an age of 2.5 Gyr in order to achieve closed orbit and provide reliable estimates of the Galactic orbit parameters for each cluster. The calculated orbital parameters are listed in Table~\ref{tab:Final_table}: $R_{\rm a}$ and $R_{\rm p}$ are apogalactic and perigalactic distances, while $e$ and $Z_{\rm max}$ are eccentricity of the orbit and the maximum vertical distance from Galactic plane respectively. $U, V, W$ and $P_{\rm orb}$ represent space velocity components and orbital period, respectively.  

The space velocity components ($U, V, W$) were corrected to the velocity components of the local standard of rest (LSR) by using values of \citet{Coskunoglu11}. These values are  given as ($8.83\pm0.24$, $14.19\pm0.34$, $6.57\pm0.21$) km s$^{-1}$. The LSR corrected space velocity components were estimated $(U, V, W)_{\rm LSR}$ = ($18.92\pm3.05$, $-11.44\pm1.41$, $8.58\pm0.32$) km s$^{-1}$ for King 6, and $(U, V, W)_{\rm LSR}$ = ($12.19\pm4.72$, $-19.77\pm1.48$, $-2.64\pm6.47$) km s$^{-1}$ for NGC 1605. From these values, the total space velocity is estimated to be $S_{\rm LSR}=23.72\pm3.38$ km s$^{-1}$ for King 6 and $S_{\rm LSR}=23.38\pm8.15$ km s$^{-1}$ for NGC 1605 (see also Table~\ref{tab:Final_table}). The $S_{\rm LSR}$ results show that both of the clusters belong to the young thin-disc stars \citep{Leggett92}. Moreover, the maximum vertical distance from the Galactic plane and eccentricity of King 6 ($Z_{\rm max}=142\pm3$ pc, $e=0.066\pm0.012$) and NGC 1605 ($Z_{\rm max}=68\pm5$ pc, $e=0.076\pm0.018$) imply that the two clusters are members of the thin-disc component of the Galaxy and move in nearly circular orbits around the Galactic centre. 

\begin{figure*}
\centering
\includegraphics[scale=0.28, angle=0]{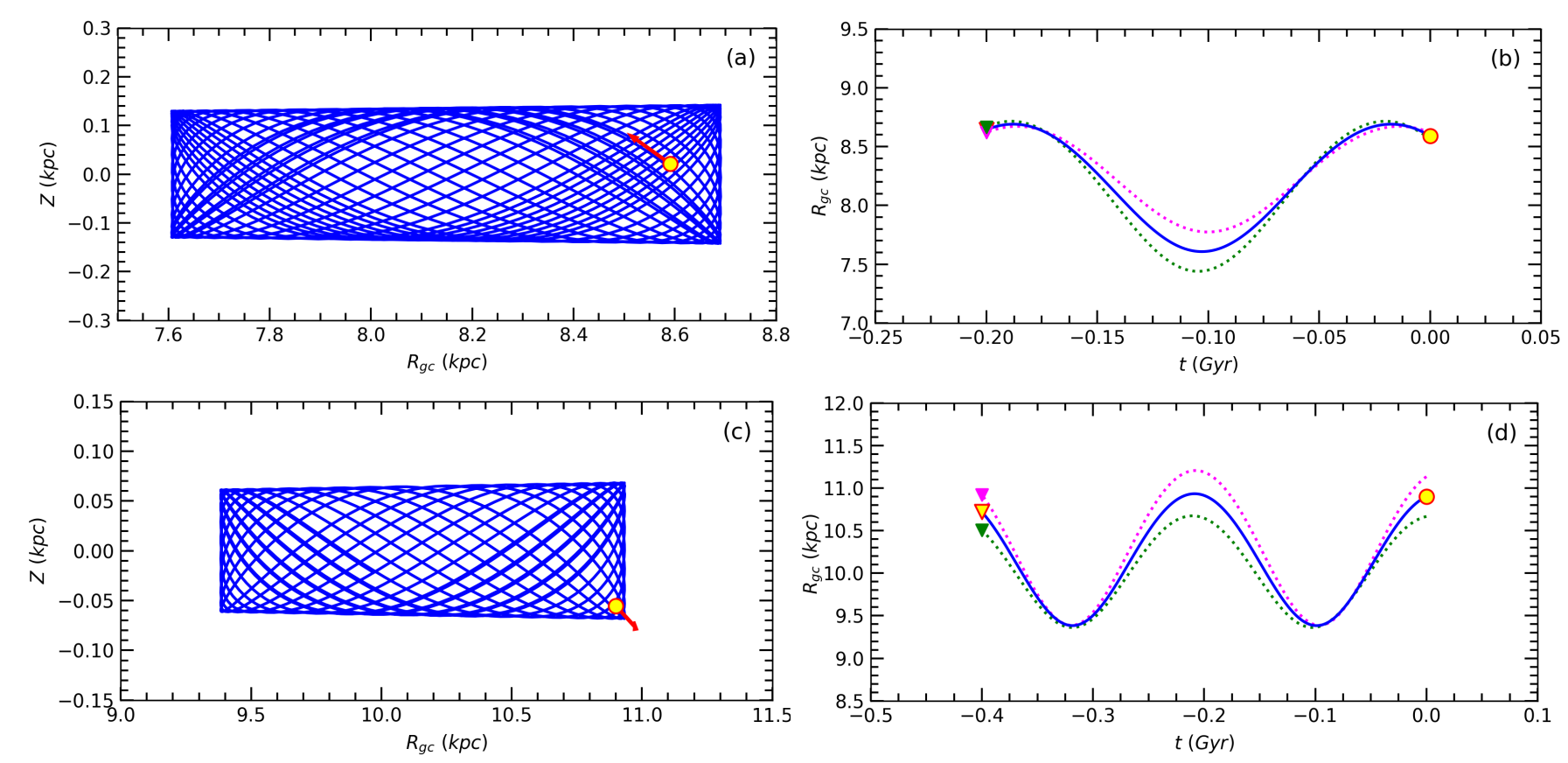}
\caption{\label{fig:galactic_orbits}
The Galactic orbit of King 6 (panels a-b) and NGC 1605 (panels c-d) in the $Z \times R_{\rm GC}$ and $R_{\rm GC} \times t$. The filled yellow circles are the present-day positions of the clusters.  The triangle symbols indicate the birth positions. The red arrows show the motion vectors of the clusters. The pink and green dotted lines show the orbit when errors in input parameters are considered.  The pink-filled triangles represent the birth locations of the cluster 
 based on the upper error estimates, while the green-filled triangles represent the birth locations based on the lower error estimates.} 
\end {figure*}

Fig.~\ref{fig:galactic_orbits} represents the orbits of the King 6 and NGC 1605. Figure~\ref{fig:galactic_orbits}a (King 6) and Figure~\ref{fig:galactic_orbits}c (NGC 1605) show the movement of the clusters on $Z \times R_{\rm GC}$ as the `side view' orbits, where the red arrows indicate the directions of motion for the clusters. Figure~\ref{fig:galactic_orbits}b (King 6) and Figure~\ref{fig:galactic_orbits}d (NGC 1605) represent the clusters' changing distance from the Galactic centre with time on the $R_{\rm GC} \times t$ plane. Results show that the orbits of King 6 and NGC 1605 follow a `boxy' orbital pattern in the Galaxy. The present-day location of the clusters are marked with yellow-filled circles, while the birth radii of the two clusters are given by the yellow filled triangles (Figs.~\ref{fig:galactic_orbits}b and \ref{fig:galactic_orbits}d). The birth radius for King 6 was estimated as $8.64\pm0.05$ kpc and that of NGC 1605 as $10.72\pm0.44$ kpc. These values show that both of the clusters formed outside the solar circle. Considering the $R_{\rm p}$ and $R_{\rm a}$ distances we infer that despite the formation location, King 6 crosses the solar circle during its orbital movement (see Fig.~\ref{fig:galactic_orbits}a), while NGC 1605 completely orbits outside the solar circle (see Fig.~\ref{fig:galactic_orbits}c). Birth radii of the two clusters were also calculated by taking into account the uncertainties in proper motion, radial velocity, and distance. The pink and green dotted lines in Figs.~\ref{fig:galactic_orbits}b (King 6) and \ref{fig:galactic_orbits}d (NGC 1605) illustrate the cluster movement in time when upper and lower errors of input parameters are adopted.

\section{Dynamical Study of the Clusters}

\subsection{Luminosity Functions}
\label{sec:LF}

\begin{figure}
\centering
\includegraphics[scale=1, angle=0]{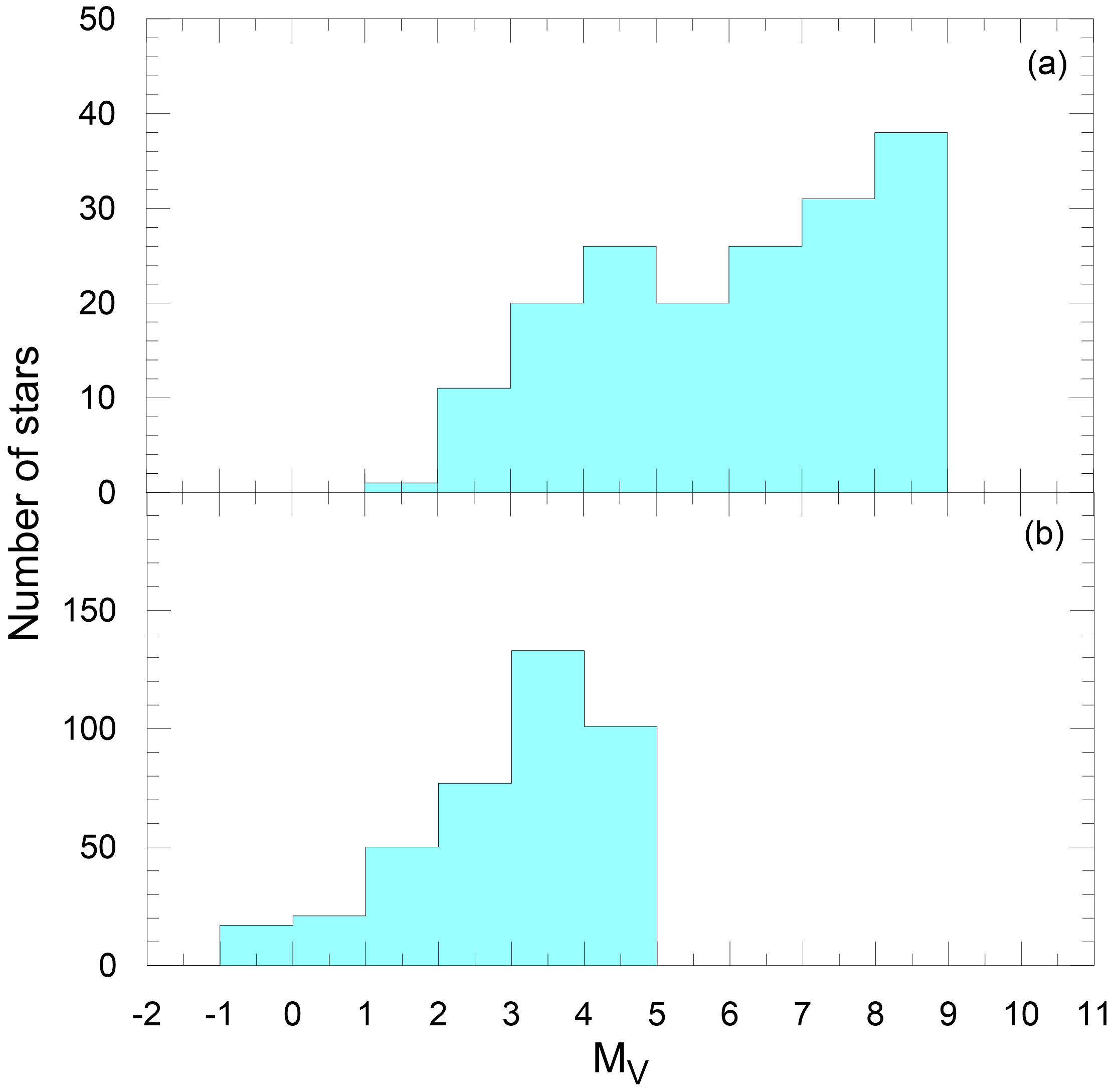}
\caption{\label{fig:luminosity_functions}
The luminosity functions for King 6 (a) and NGC 1605 (b) open clusters.}
\end {figure}

We constructed the luminosity function (LF) for each cluster through consideration of the main-sequence stars with membership probabilities $P>0$ and positions within the limiting radii of the clusters ($r_{\rm lim}^{\rm obs}\leq10'$). In order to obtain more precise information about the  LFs of the clusters and the related mass functions, it is important to consider stars with membership probabilities greater than 0 in the calculations \citep{Akbulut21}. We transformed apparent $V$ magnitudes to their absolute $M_{\rm V}$ magnitudes using the distance modulus definition $M_{\rm V} = V-5\times \log d + 5 - 3.1\times E(B-V)$, where $V$, $d$ and $E(B-V)$ indicate apparent magnitude, isochrone distance, and color excess as previously obtained for the two clusters (Table~\ref{tab:Final_table}). The absolute magnitude range of the stars is $1.22< M_{\rm V}< 8.98$ mag and $-0.90< M_{\rm V}< 4.98$ mag for King 6 and NGC 1605, respectively. Despite both of the clusters being of relatively young age, as King 6 is closer to the Sun than NGC 1605 its absolute magnitude range is wider. The histograms of LF with the intervals of 1 mag are given in Fig.~\ref{fig:luminosity_functions}. It can be interpreted from the figure that the LF of King 6 reaches to approximately $M_{\rm V} = 9$ mag (Fig.~\ref{fig:luminosity_functions}a), and that this limit is $M_{\rm V}=5$ mag for NGC 1605 (Fig.~\ref{fig:luminosity_functions}b).  

\newpage
\subsection{Mass Functions}
\label{sec:MF}

The mass function (MF) expression that we used is described as follows:
\begin{eqnarray}
{\rm log(dN/dM)}=-(1+\Gamma)\times \log M + {\rm constant}.
\label{eq:mass_luminosity}
\end{eqnarray}
Here $dN$ is the number of stars per unit mass range $dM$, the central mass is designated as $M$, and $\Gamma$ is the slope of the MF.  

\begin{figure}
\centering
\includegraphics[scale=1, angle=0]{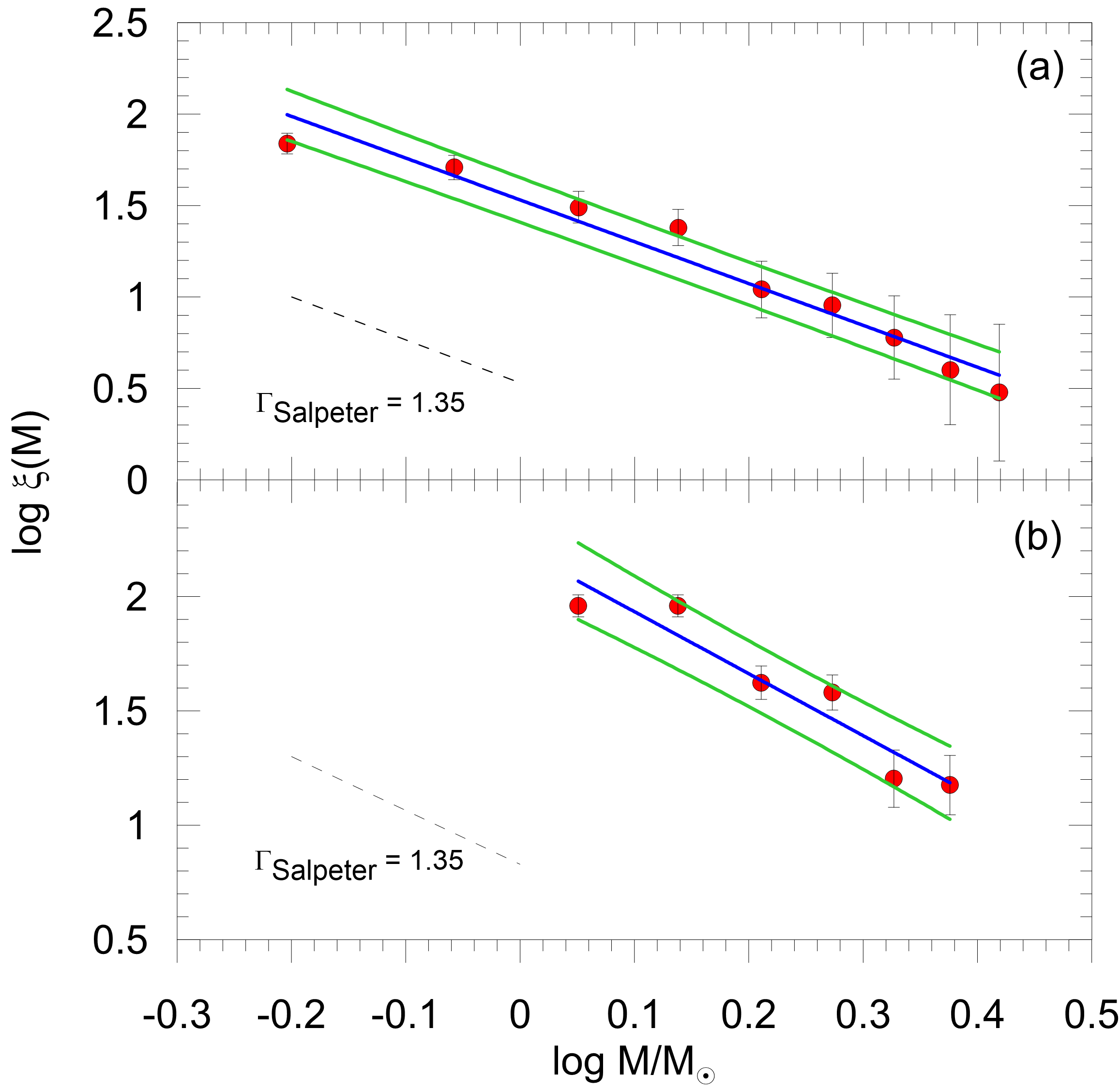}
\caption{\label{fig:mass_functions}
Mass functions of King 6 (a) and NGC 1605 (b). Blue lines show the mass functions of the open clusters, while green lines are the $\pm1\sigma$ standard deviations. The grey dashed-lines in the panels represent the slope of \citet{Salpeter55}.}
\end {figure}

To derive MFs of the studied clusters, we used {\sc parsec} isochrones from \citet{Bressan12} that matched the metallicity fractions ($z$) as estimated in the current study. We generated a high-degree polynomial equation between the theoretical $V$-band absolute magnitudes and masses from the selected isochrones. Through the application of this equation, we transformed the observational absolute magnitudes $M_{\rm V}$ of the stars to the their masses. The membership probabilities of stars used in the MF analyses are $P>0$. There are 213 such stars in King 6, and 462 in NGC 1605. These correspond to stellar masses within the $0.58\leq M/ M_{\odot}\leq 3.59$ and $1.03\leq M/ M_{\odot}\leq 2.94$ for the relevant clusters, respectively. The MF slope values were derived as $\Gamma=1.29 \pm 0.18$ for King 6 and as $\Gamma=1.63 \pm 0.36$ for NGC 1605. These results agrees with the \citet{Salpeter55}'s value of 1.35 within the uncertainties. The MF slopes for the two clusters were plotted in Fig.~\ref{fig:mass_functions} . Considering the stellar mass ranges used in MF estimation, we estimated the total cluster masses as $ 195 M/M_{\odot}$ and $623 M/M_{\odot}$ for King 6 and NGC 1605, respectively.

\begin{figure}
\centering
\includegraphics[scale=0.80, angle=0]{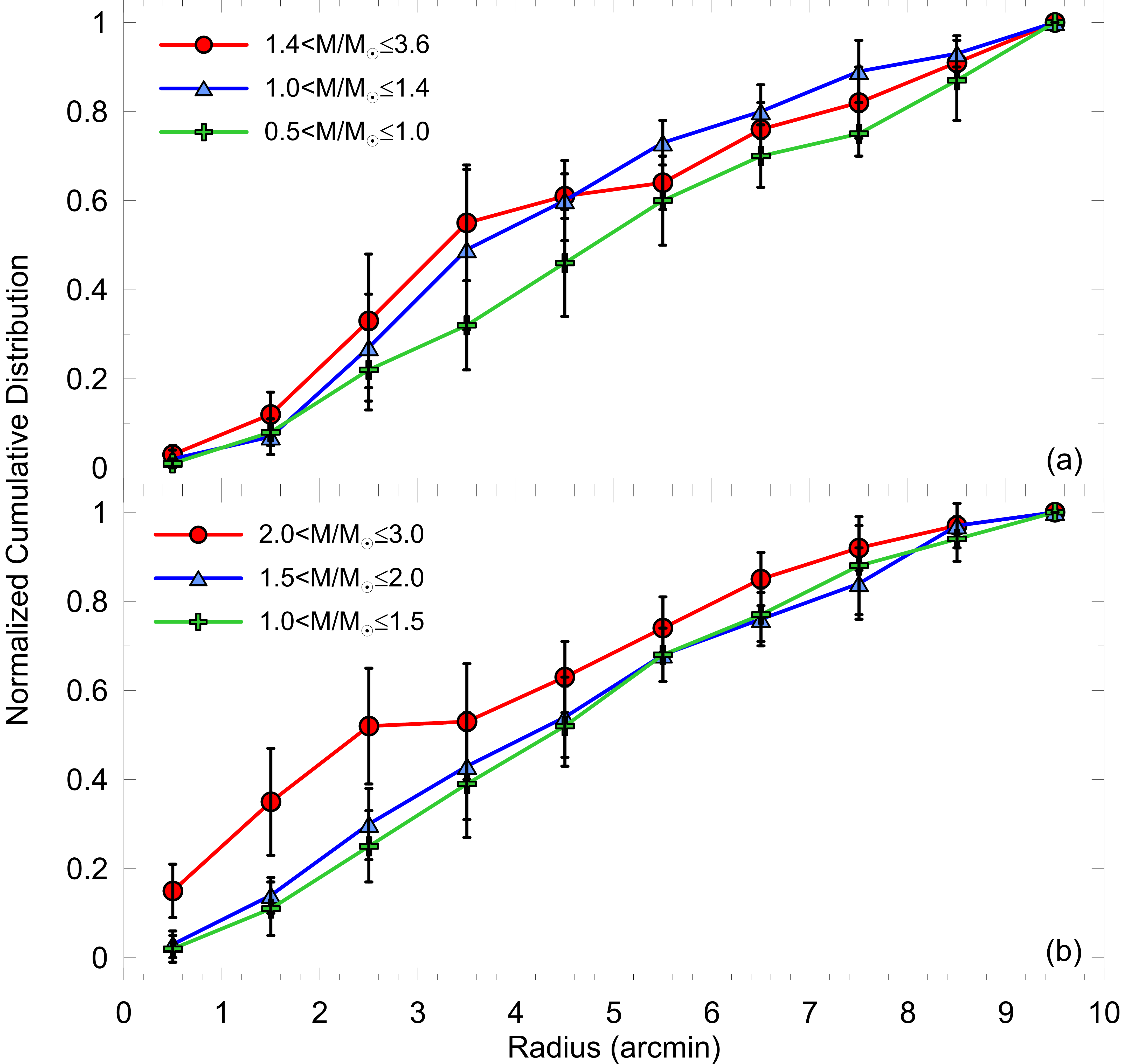}
\caption{\label{fig:radial_distributions}
The cumulative radial distribution of stars in different mass ranges for King 6 (a) and NGC 1605 (b).}
\end {figure}

\subsection{The Dynamical State of Mass Segregation}

Mass segregation could have significant association with the dynamical evolution and lifetime of open clusters. Several mechanisms such as primordial mass segregation, two-body relaxation, mass-dependent stellar evolution, stellar dynamics and mass segregation feedback might contribute to mass segregation in OCs  \citep{Raboud98, Fischer98, Alcock19, Sariya21, Piecka21}. Primordial mass segregation refers to dynamical interactions and gravitational collapse during the formation of an OC that makes massive stars sink towards the central regions \citep{deGrijs03, Pavlik20}. Over time gravitational interactions between cluster members lead to a process of two-body relaxation which causes stars to exchange kinetic energy and momentum, leading to massive stars moving towards the cluster center \citep{Sagar88, delaFuente96, Bisht20, Pavlik22}. Moreover, the mass segregation itself can influence the dynamics of the cluster. The more concentrated distribution of massive stars in the cluster center enhances the interaction rate between them, leading to more frequent gravitational encounters. These encounters might force core-collapse processes, such as core-collapse-induced star formation or binary formation, extending the mass segregation \citep{Pang13}.

Related to mass segregation, the relaxation time indicates the timescale for two-body relaxation processes to occur within the cluster. The transmission of the energy occurs between massive stars to low-mass stars, leading the stellar velocity distribution to be Maxwellian \citep[see e.g.,][]{Hillenbrand98, Baumgardt03, Dib19}. The relaxation time, denoted as $T_{\rm E}$, can be estimated using the following formula given by \citet{Spitzer71}:
\begin{equation}
T_{\rm E} = \frac{8.9 \times 10^{5} N^{1/2} \times R_{\rm h}^{3/2}}{\langle m\rangle^{1/2}\times\log(0.4N)},
\end{equation} 
where $N$ is the total number of stars, $R_{\rm h}$\footnote{The half-mass radius of the cluster is the radius which contains half the total mass in relevant cluster.} is the half-mass radius in parsecs, and $\langle m \rangle$ is the average mass of the considered stars in solar units. 

To calculate relaxation times for the studied clusters, we used the stars with membership probabilities $P>0$, located within the cluster limiting radii ($r_{\rm lim}^{\rm obs}\le10'$) and brighter than photometric completeness limit ($V\leq20$ mag), which were mentioned in Sections~\ref{sec:LF} and \ref{sec:MF}. The calculated mean stellar mass is $\langle m \rangle = 1.06 M/M_{\odot}$ for King 6 and $\langle m \rangle = 1.55 M/M_{\odot}$ for NGC 1605. Half-mass radii are  $R_{\rm h}$= 0.95 pc for King 6 and $R_{\rm h}$= 4.01 pc for NGC 1605. The dynamical relaxation time of King 6 was estimated as $T_{\rm E}=5.8$ Myr and for NGC 1605 as $T_{\rm E}=60$ Myr. We conclude that both clusters are dynamically relaxed due to the derived $T_{\rm E}$ ages being younger than the present ages of the two clusters as estimated in the current study (see Table~\ref{tab:Final_table}).

To understand the impact of the mass segregation effect in the two clusters, we divided the masses of selected stars into three intervals. These ranges are $0.5 < M/M_{\odot}\leq 1$ (low-mass), $1 < M/M_{\odot}\leq 1.4$ (intermediate-mass), and $1.4 < M/M_{\odot}\leq 3.6$ (high-mass) for King 6 and $1 < M/M_{\odot}\leq 1.5$, $1.5 < M/M_{\odot}\leq 2$, and $2 < M/M_{\odot}\leq 3$ for NGC 1605. The normalized cumulative radial distributions of stars in these different mass ranges are shown in Fig.~\ref{fig:radial_distributions}. Generally, the normalized cumulative radial distributions of stars in both of the open clusters represent a mass segregation effect as bright stars seem to be more centrally concentrated than the low-mass members. We used the Kolmogorov–Smirnov test, finding the confidence level for a mass segregation effect to be 91\% for both open clusters. 

\begin{table}
  \centering
  \renewcommand{\arraystretch}{0.8}
  \caption{Fundamental parameters of King 6 and NGC 1605.}
  \medskip
  {\small
        \begin{tabular}{lrr}
\hline
Parameter & King 6 & NGC 1605 \\
\hline
($\alpha,~\delta)_{\rm J2000}$ (Sexagesimal)& 03:27:55.67, $+$56:26:38.40 & 04:34:58.80, $+$45:16:08.40\\
($l, b)_{\rm J2000}$ (Decimal)              & 143.3444, $-$0.0949         & 158.5860, $-01$.5673       \\
$f_{0}$ (stars arcmin$^{-2}$)               & 2.28 $\pm$ 0.24             & 13.05 $\pm$ 0.73           \\
$f_{\rm bg}$ (stars arcmin$^{-2}$)          & 5.12 $\pm$ 0.16             & 9.81 $\pm$ 0.30            \\
$r_{\rm c}$ (arcmin)                        & 4.68 $\pm$ 1.07             & 1.90 $\pm$ 0.20            \\
$r_{\rm lim}^{\rm obs}$ (arcmin)            & 10                          & 10                         \\
$r$ (pc)                                    & 2.10                        & 8.88                       \\
Cluster members ($P\geq0.5$)                & 112                         & 160                        \\
$\mu_{\alpha}\cos \delta$ (mas yr$^{-1}$)   & +3.833 $\pm$ 0.034          & +0.928 $\pm$ 0.104         \\
$\mu_{\delta}$ (mas yr$^{-1}$)              & $-1.906 \, \pm$ 0.032       & $-1.997 \, \pm$ 0.082      \\
$\varpi$ (mas)                              & 1.381 $\pm$ 0.042           & 0.336 $\pm$ 0.043          \\
$d_{\varpi}$ (pc)                           & 724 $\pm$ 22                & 2976 $\pm$ 381             \\
$E(B-V)$ (mag)                              & 0.515 $\pm$ 0.030           & 0.840 $\pm$ 0.054          \\
$E(U-B)$ (mag)                              & 0.371 $\pm$ 0.022           & 0.605 $\pm$ 0.039          \\
$A_{\rm V}$ (mag)                           & 1.596 $\pm$ 0.093           & 2.604 $\pm$ 0.167          \\
$[{\rm Fe/H}]$ (dex)                        & 0.02 $\pm$ 0.20             & 0.01  $\pm$ 0.20     \\
Age (Myr)                                   & 200 $\pm$ 20                & 400 $\pm$ 50               \\
$V-M_{\rm V}$ (mag)                         & 10.892 $\pm$ 0.099          & 15.028 $\pm$ 0.167         \\
Isochrone distance (pc)                     & 723 $\pm$ 34                & 3054 $\pm$ 243             \\
$(X, Y, Z)_{\odot}$ (pc)                    & ($-580$, 432, 1)            & ($-2842$, 1115, 84)        \\
$R_{\rm GC}$ (kpc)                          & 8.59                        & 10.90                      \\
MF slope                                    & 1.29 $\pm$ 0.18             & 1.63 $\pm$ 0.36            \\
Total mass ($M/M_{\odot}$)                  & 195                         & 623                        \\ 
$V_{\rm R}$ (km/s)                          & $-$23.40 $\pm$ 3.26         &  $-$15.27 $\pm$ 1.35       \\
$U_{\rm LSR}$ (km/s)                        & 18.92 $\pm$ 3.05            &  12.19 $\pm$ 4.72          \\
$V_{\rm LSR}$ (km/s)                        & $-$11.44 $\pm$ 1.41         &  $-$19.77 $\pm$ 1.48       \\
$W_{\rm LSR}$ (km/s)                        & 8.58 $\pm$ 0.32             &  $-$2.64 $\pm$ 6.47        \\
$S_{_{\rm LSR}}$ (km/s)                     & 23.72 $\pm$ 3.38            &  23.38 $\pm$ 8.15          \\
$R_{\rm a}$ (pc)                            & 8691 $\pm$ 22               &  10933 $\pm$ 98            \\
$R_{\rm p}$ (pc)                            & 7607 $\pm$ 168              &  9384 $\pm$ 50             \\
$Z_{\rm max}$ (pc)                          & 142 $\pm$ 3                 &  68 $\pm$ 5                \\
$e$                                         & 0.066 $\pm$ 0.012           &  0.076 $\pm$ 0.018         \\
$P_{\rm orb}$ (Myr)                         & 228 $\pm$ 2                 &  292 $\pm$ 1               \\
Birthplace (kpc)                            & 8.64 $\pm$ 0.05             &  10.72 $\pm$ 0.44          \\
\hline
        \end{tabular}%
    } 
    \label{tab:Final_table}%
\end{table}%


\section{Summary and Conclusion}

We investigated two open clusters, King 6 and NGC 1605, which are located in the second Galactic quadrant, using newly acquired CCD {\it UBV} and {\it Gaia} DR3 data. The membership probabilities of stars were calculated in a 5-dimensional spatial distribution where {\it Gaia} astrometric data and their uncertainties were taken into consideration. In addition, we limited the star selection according to the limiting radius, ZAMS fits, and the completeness limit of each cluster. Thus, we end up with 112 and 160 `most likely' member stars with membership probabilities $P\geq0.5$ for King 6 and NGC 1605,  respectively. We used these stars during the subsequent estimation of the astrophysical parameters of the two clusters. 
\citet{Ann02} indicated that the reddening in the direction of King 6 can not be obtained as a single value due to the differential reddening affect, they also noted the possible binary sequence above the cluster's ZAMS. In this study, when {\em UBV} and {\it Gaia} based CMDs and positions of the most probable main-sequence member stars on TCDs were investigated, differential reddening effects were not found. In addition, the binary star sequence above the ZAMS mentioned by \citet{Ann02} was not clearly detected for King 6 in our study. We concluded that this may be due to the precision of the photometric data.

Based on the investigation in the RDP of NGC 1605, \citet{Camargo21} detected an increase in the number density of stars and suggested the possibility of a second cluster (i.e., NGC 1606 is a binary cluster). \citet{Camargo21} also found a large difference (about 1.4 Gyr) in age between two clusters and mentioned that this was due to effects of tidal capture during a close encounter of the two clusters. In this study, the RDP of NGC 1605 shows a small increase in the stellar density between 7 and 9 arcmin from the cluster centre (see Fig~3b), but direct evidence for a second cluster was not found. Examination of the radial velocities of faint stars does not support existence of a second cluster, as shown above. \citet{Anders22}, who recently analysed the open cluster NGC 1605 with {\it Gaia} EDR3 data \citep{Gaia21}, also found no evidence that the cluster is a pair.

A summary of the main findings of the study is listed as follows:  

\begin{enumerate}

    \item{Considering the best solution of the  RDP fitting, we estimated limiting radii by visual inspection of the data for the two clusters. These values are $r_{\rm lim}^{\rm obs}=10'$ for both King 6 and NGC 1605, which are compatible with the calculated limiting radii for each cluster.}

    \item{VPD analyses showed that the member stars of King 6 are clearly separated from the field stars, whereas member stars of NGC 1605 nest together with the background stars. Mean proper motion values were estimated as ($\mu_{\alpha}\cos \delta, \mu_{\delta})=(3.833\pm 0.034, -1.906\pm 0.032$) mas yr$^{-1}$ for King 6 and as ($\mu_{\alpha}\cos \delta, \mu_{\delta}) = (0.928\pm 0.104, -1.997\pm 0.082$) mas yr$^{-1}$ for NGC 1605. The mean trigonometric parallaxes were derived as $\varpi = 1.381 \pm 0.042$ mas for King 6 and $\varpi = 0.336 \pm 0.043$ mas for NGC 1605. These equate to distances of $d_{\rm \varpi}=724\pm 22$ pc and  $d_{\rm \varpi}=2976\pm 381$ pc, respectively.}

    \item{Color excesses and photometric metallicities of the two clusters were obtained individually from observational $(U-B)\times (B-V)$ TCDs constructed from the `most likely' (as defined above) cluster main-sequence stars.  Analyses provided the color excesses to be $E(B-V)=0.515\pm 0.030$ mag for King 6 and $E(B-V)=0.840\pm 0.054$ mag for NGC 1605, photometric metallicities to be [Fe/H]$= 0.02\pm 0.20$ dex for King 6 and [Fe/H]$= 0.01\pm 0.20$ dex for NGC 1605.}

    \item{We constructed CMDs from the most likely cluster members for {\it UBV} and {\it Gaia} DR3 photometric data in order to derive distance moduli and ages. Keeping color excesses and metallicity as constants, we fitted {\sc parsec} isochrones scaled by the estimated $z$ values to the CMDs. The apparent distance modulus, distance, and age of King 6 were estimated to be $\mu_{\rm V}=10.892\pm 0.099$ mag, $d_{\rm iso}=723 \pm 34$ pc, and $t=200 \pm 20$ Myr, respectively. These values correspond to $\mu_{\rm V}=15.028\pm 0.167$ mag, $d_{\rm iso}=3054 \pm 243$ pc, and $t=400 \pm 50$ Myr for NGC 1605. The CMD-based distances are in agreement, within the errors, with the distances derived by taking into account the trigonometric parallaxes ($d_{\rm \varpi}$).}

    \item{Radial velocities for member stars with membership probabilities $P\geq0.8$ were taken from {\it Gaia} DR3 database to calculate the clusters' mean radial velocity values and to investigate their Galactic orbital parameters. Thus, for King 6 we used 23 stars and derived a mean $V_{\rm R} = -23.40\pm 3.26$ km~s$^{-1}$. For NGC 1605 we used one star with the value of $V_{\rm R} = -15.27\pm 1.35$ km ~s$^{-1}$. Orbit integrations showed that both of the clusters belong to the young thin-disc population of the Galaxy and formed outside the solar circle.}

    \item{The mass function slopes of King 6 and NGC 1605 were calculated as $\Gamma=1.29\pm 0.18$ and $\Gamma=1.63\pm 0.36$, respectively. These findings are compatible with the value of 1.35 given by \citet{Salpeter55}.}
    
    \item{Mass segregation is observed in both of the OCs. The K-S test indicates at the 91\% confidence level that this effect is present for both of the clusters. The dynamical relaxation times are less than both of the OC ages, demonstrating that King 6 and NGC 1605 are dynamically relaxed.}
    
\end{enumerate}

\newpage
\software{IRAF \citep{Tody86, Tody93}, PyRAF \citep{Science12}, SExtractor \citep{Bertin96}, Astrometry.net \citep{Lang10}, UPMASK \citep{Krone-Martins14}, pyUPMASK \citep{Pera21}, GALPY \citep{Bovy15}, MWPotential2014 \citep{Bovy15}.}


\acknowledgments
This study is partly based on Sevinc Gokmen’s MSc thesis. This study has been supported in part by the Scientific and Technological Research Council (T\"UB\.ITAK) 113F270 and 122F109. This study was funded by Scientific Research Projects Coordination Unit of Istanbul University. Project number: 39743. We thank T\"UB\.ITAK for partial support towards using the T100 telescope via project 18CT100-1396. We also thank the on-duty observers and members of the technical staff at the T\"UB\.ITAK National Observatory for their support before and during the observations. We thank Remziye Canbay and Seval Ta{\c{s}}demir for their contribution to data reduction and literature review studies. This research has made use of the WEBDA database, operated at the Department of Theoretical Physics and Astrophysics of the Masaryk University. We also made use of VizieR and Simbad databases at CDS, Strasbourg, France. Data were sourced from the European Space Agency (ESA) mission \emph{Gaia}\footnote{https://www.cosmos.esa.int/gaia}, processed by the \emph{Gaia} Data Processing and Analysis Consortium (DPAC)\footnote{https://www.cosmos.esa.int/web/gaia/dpac/consortium}. Funding for DPAC has been provided by national institutions, in particular the institutions participating in the \emph{Gaia} Multilateral Agreement. We are grateful for the analysis system IRAF, which was distributed by the National Optical Astronomy Observatory (NOAO). NOAO was operated by the Association of Universities for Research in Astronomy (AURA) under a cooperative agreement with the National Science Foundation. PyRAF is a product of the Space Telescope Science Institute, which is operated by AURA for NASA. We thank the anonymous referee for their helpful comments and guidance which helped improve this paper.


\end{document}